\documentclass[twocolumn,showpacs,preprintnumbers,amsmath,amssymb]{revtex4}

\usepackage{graphicx}
\usepackage{dcolumn}
\usepackage{bm}

\def\bea{\begin{eqnarray}}
\def\eea{\end{eqnarray}}

\def\pp{\mbox{$p$-$p$} }
\def\auau{\mbox{Au-Au} }

\def\aa{\mbox{A-A} }
\def\nn{\mbox{N-N} }

\begin{document} 

\preprint{Version 1.6}

\title{Dijet production, collision centrality and backgrounds in high-energy p-p collisions
}

\author{Thomas A.\ Trainor}
\address{CENPA 354290, University of Washington, Seattle, WA 98195}


\date{\today}

\begin{abstract}
{\bf Background:} Two aspects of high-energy \pp collisions share common phenomenological elements: 
(a) A correlation between jet production and \pp centrality is suggested by the transverse partonic structure of hadrons inferred from deep-inelastic scattering data. 
(b) The {\em underlying event} (UE) is defined as the final-state particles complementary to a triggered high-energy dijet. 
An observable common to both topics is variation of so-called {\em transverse multiplicity} $N_\perp$ with a $p_{t,trig}$ dijet trigger.
{\bf Purpose:}
Test assumptions associated with \pp collision centrality and the UE. Determine the nature of the UE and explore the relation between jet production and \pp centrality.
{\bf Method:} 
Use the {\em two-component model} (TCM) of spectra and correlations derived from 200 GeV \pp collisions to construct a simulated particle distribution on $(p_t,n_{ch})$. Use the simulation to predict the $N_\perp$ response to $p_{t,trig}$. Use the measured  $p_t$ spectrum $n_{ch}$ dependence to explore the effect of changing \pp centrality.
{\bf Results:} 
The TCM provides a good description of $N_\perp$ vs $p_{t,trig}$ and the $N_\perp(p_t)$ spectrum. The relation of $N_\perp$ to $p_t$ and $p_{t,trig}$ departs from assumptions.
{\bf Conclusions:}
The $p_t$ spectrum TCM combined in this analysis with measured minimum-bias \pp angular correlations suggests that the UE includes a substantial contribution from the triggered dijet in addition to the contribution from projectile fragmentation (beam-beam remnants). 
The jet contribution to $N_\perp$ may represent a universal large-angle base common to all dijets that extends across $2\pi$ azimuth. The analysis further suggests that \pp centrality is not controlled significantly by $p_{t,trig}$ but may be correlated to some extent with an imposed $n_{ch}$ condition, depending on the role of fluctuations. Future correlation studies may better determine the role of \pp centrality. These results may have implications for ongoing RHIC analysis and LHC searches for physics beyond the standard model.
\end{abstract}

\pacs{12.38.Qk, 13.87.Fh, 25.75.Ag, 25.75.Bh, 25.75.Ld, 25.75.Nq}

\maketitle

 \section{Introduction}

The hadronic final state of high-energy \pp collisions is of fundamental importance as a reference for interpretation of nucleus-nucleus collisions at the RHIC and LHC. The search for novel physics in any new collision system or energy regime must refer to what is ``conventional'' in a standard-model QCD description of elementary hadronic collisions. There are several possible contributions to single-particle spectra and two-particle correlations, including transverse fragmentation of large-angle-scattered partons to jets, longitudinal projectile-nucleon dissociation, radial and other flow components, Bose-Einstein correlations (BEC) and possibly other processes not yet anticipated theoretically.

Of particular interest are two possibly-related issues: (a) Are there significant manifestations of \pp collision centrality in final-state hadron spectra and correlations? Is the concept of collision centrality viable in a small system possibly dominated by fluctuations? If so, how is \pp centrality related to final-state observables, especially jet production? (b) What is the nature of the {\em underlying event} (UE), those \pp collision products characterized by {\em transverse multiplicity} $N_\perp$ (relative to the dijet axis) that are assumed to be distinct from a triggered high-energy dijet? What physical mechanisms may contribute? 

In Ref.~\cite{ppcent2} deep-inelastic scattering (DIS) data are interpreted to suggest that the low-$x$ gluon component of the proton wave function is confined to a radius substantially smaller than the overall hadron size. Jet production at mid rapidity in \pp collisions might then correspond  to reduced impact parameter. A $p_t$ trigger condition $p_{t,trig}$ should favor jet production at higher $p_t$. A nonjet component of the event multiplicity, transverse multiplicity $N_\perp$, could be a measure of collision centrality. 
A specific trend of $N_\perp$ vs  $p_{t,trig}$ might then confirm the conjectured centrality dependence and identify the $p_{t,trig}$ interval favoring jet production.

A high-$p_t$ trigger condition may serve as a form of jet selection where event-wise jet reconstruction is impractical.  But other methods are available to isolate or infer jet systematics, including differential spectrum analysis~\cite{ppprd,hardspec} and combinatoric two-particle correlation analysis~\cite{porter2,porter3}. In Ref.~\cite{ppprd} a two-component (soft+hard) analysis of $p_t$ spectra for several \pp event-multiplicity classes revealed a spectrum ``hard'' component that scales with multiplicity as $n_{ch}^2$ and has properties consistent with a {\em minimum-bias} jet contribution (no condition on jet energy). The ``soft'' component scales linearly with $n_{ch}$ and is consistent with projectile-proton dissociation (beam-beam remnants). The relation of the hard component to perturbative QCD (pQCD) predictions was established in Ref.~\cite{fragevo}. Two-particle correlations from \pp collisions have also been related to expected jet systematics~\cite{porter2,porter3}.

In the present study we consider the proposal from Ref.~\cite{ppcent2} in the context of  the $p_t$ spectrum systematics described in Ref.~\cite{ppprd}. We use spectrum information to generate jet-related (hard) and nonjet (soft) contributions to hadron production distributed on the 2D space $(p_t,n_{ch})$. We then determine the effect of a $p_{t,trig}$ condition on  soft and hard components compared to expectations from Ref.~\cite{ppcent2}. The present analysis could be extended to jet structure in  2D angular correlations~\cite{porter2,porter3,anomalous} and a {\em nonjet quadrupole} component observed to depend strongly on collision geometry in \aa collisions~\cite{davidhq,davidhq2}.

This study also addresses the {\em underlying event}.  Variation of UE properties (azimuth correlation structure and $N_\perp$) with event selection based on a $p_{t,trig}$ condition has been studied extensively~\cite{rick,under5,under}. There is a formal relation with the proposal of Ref.~\cite{ppcent2} in terms of $N_\perp$ response to variation of $p_{t,trig}$. In both cases $N_\perp$ is assumed to be unrelated to the triggered dijet and sensitive to the properties of the underlying event.  In the present analysis we combine the event-type selection model developed in relation to Ref.~\cite{ppcent2} with two-component spectrum and angular-correlation phenomenology from 200 GeV \pp collisions~\cite{porter2,porter3} to develop an alternative interpretation of $N_\perp$ systematics and the nature of the UE.

Finally, the measured $n_{ch}$ systematics of jet production in spectra and correlations from 200 GeV \pp collisions are compared with a pQCD description of dijet production in the context of a Glauber model based on transverse parton distributions within projectile nucleons. We seek evidence from the trend of jet production vs $n_{ch}$ for some relation to \pp centrality. The event-wise $n_{ch}$ condition is complementary to the $p_{t,trig}$ condition employed in UE studies and proposed in Ref.~\cite{ppcent2} to demonstrate the relevance of \pp centrality to jet production. 

This paper is arranged as follows: 
In Sec.~\ref{basic} we introduce two descriptions of jet production in \pp collisions in the context of proton transverse structure, \pp centrality and the underlying event that provide a context for the present study. 
In Sec.~\ref{twocomp} we review a two-component model of particle yields, spectra and correlations inferred from 200 GeV \pp collision data. 
In Sec.~\ref{cdftrans} we analyze the structure of the transverse multiplicity spectrum $dN_\perp(p_t)/dp_t$ for triggered events at 0.9 and 1.8 TeV.
In Sec.~\ref{minbias} we apply the two-component model to the \pp minimum-bias event distribution on $n_{ch}$. 

In Sec.~\ref{ttotal} we construct ensemble-averaged particle densities on event multiplicity $n_{ch}$ and transverse rapidity $y_t$ for soft and hard  hadron production mechanisms and
%
for soft (no jets) and hard (at least one jet) event types.
In Sec.~\ref{angcorr} we present measured angular correlations for minimum-bias jets in 200 GeV \pp collisions, emphasizing their contribution to the transverse azimuth region and $N_\perp$.
In Sec.~\ref{transverse} we illustrate the relation between transverse multiplicity $N_\perp$ and trigger $p_t$ cut that is a central element of underlying-event studies.
In Sec.~\ref{gglauber} we review the Glauber model applied to \auau and \pp collisions and its implications for a correlation between particle and dijet production and \pp collision centrality.
In Sec.~\ref{disc} we compare the results of this study to the material presented in Sec.~\ref{basic}. Sec.~\ref{summ} summarizes.

\section{Basic structure of $\bf p$-$\bf p$ collisions} \label{basic}

We consider descriptions of \pp dynamics and final-state structure in two contexts. A theoretical description is based on deep-inelastic scattering (DIS) data and considers possible manifestations of \pp collision centrality in certain observables. An experimental description is based on the azimuth correlation structure of events containing a triggered or reconstructed dijet, especially properties of the {\em underlying event} assumed to be complementary to the dijet. Each approach relies on a $p_{t,trig}$ condition. 

 \subsection{A theoretical context}


In Ref.~\cite{ppcent2} it is argued that hard particle production (i.e., large-angle gluon scattering to dijets) should be most probable at small \pp impact parameter because the transverse size of the low-$x$ gluon distribution in the proton inferred from DIS data is substantially smaller than the overall proton size. 
%
Soft particle production (not associated with a triggered dijet) should vary with $b$ over a large range and may serve as a centrality measure. Specifically, {\em transverse multiplicity} $N_\perp$ (perpendicular to a trigger-particle momentum and therefore to a dijet axis) may be correlated with $b$. Absent a jet trigger collisions should be dominated by soft processes at larger $b$ with $N_\perp \approx N_{NSD}$ (NSD denotes non-single-diffractive \pp collisions).

It is assumed that jet production should be correlated with smaller $b$ and therefore larger $N_\perp$. The larger multiplicity may then make it difficult to observe lower-$p_t$ jets directly, and semihard dynamics, and particle production with $p_t \approx$ few GeV is said to be not well understood. Therefore, indirect selection of hard processes (jets) is established with a $p_t$ condition denoted here by $p_{t,trig}$. 

For sufficiently high $p_{t,trig}$ 
$b$ should be relatively small and nearly independent of the trigger condition. Equivalently, transverse multiplicity $N_\perp$ should be nearly independent of $p_{t,trig}$ and substantially larger than for NSD \pp collisions dominated by soft proton dissociation at large $b$.
Reference~\cite{ppcent2} then poses the question (given the several assumptions): above what critical $p_{t,trig}$ value is hadron production dominated by ``hard'' parton-parton interactions in more-central \pp collisions? 
In the present analysis we examine those assumptions and expectations in the context of differential analysis of $p_t$ spectra vs \pp multiplicity and the correspondence with established jet phenomenology and pQCD predictions.


\subsection{An experimental context}

The CDF collaboration has emphasized study of the {underlying event} in $p$-$\bar p$ collisions~\cite{rick,under5}. In collisions that include a (triggered) hard parton scatter to a high-energy dijet the UE is defined as what is {\em not part of the dijet}, with several possible production mechanisms including the beam-beam remnant (soft component, projectile proton dissociation), multiple parton interactions (MPI, untriggered semihard parton scatters, see Ref.~\cite{mpi} for a review) and initial-state radiation.  The UE is explored in the {\em transverse region} (TR) said to be ``very sensitive'' to the UE~\cite{under5}. The TR, defined by $|\phi - \pi/2| < \pi/6$,  includes total multiplicity $N_\perp$ within some $\eta$ acceptance $\Delta \eta$ and is assumed to be {outside the triggered hard-scatter dijet} on azimuth. Two other complementary azimuth regions are denoted {\em toward} (including the same-side or triggered jet) and {\em away} (including the away-side or recoil jet).

CDF UE analysis is based on a particle acceptance defined by $p_t > 0.5$ GeV/c and $|\eta| < 1$. Jets are defined by a cone jet finder with radius $R < 1$ on $(\eta,\phi)$ and include jet fragment multiplicities as low as 1. Relevant issues for UE analysis include particle production and correlation mechanisms, trends of $N_\perp$ on collision energy and trigger $p_t$ condition and the $N_\perp$ $p_t$ spectrum. 


Some results from the CDF UE analysis at $\sqrt{s} = 1.8$ TeV are particularly relevant to the present study:  In Fig.~13 of Ref.~\cite{under5} the growth of total event multiplicity within the $\eta$ acceptance is plotted vs $p_{t,trig}$. In Fig.~21 the total yield is separated into three parts on azimuth relative to the trigger. The $N_\perp$ yield from the transverse region assumed to represent the UE increases with  $p_{t,trig}$ from zero to a limiting {\em plateau value} $\approx 2.3$. The increase of $N_\perp(p_{t,trig})$ is attributed to an MPI rate increasing to a saturation value. In Fig.~17 the azimuth distribution is shown for a specific {\em jet trigger} ($P_j > 5$ GeV/c) in relation to the three equal azimuth regions. And in Fig.~37 the $N_\perp$ $p_t$ spectrum for the TR is shown for three trigger conditions. The $N_\perp$ spectrum is assumed to be monolithic (described by a single functional form). In the present study we consider those results in the context of minimum-bias systematics from 200 GeV \pp collisions and a two-component model of hadron production.

\subsection{Strategy for the present analysis}

We first present two-component models for \pp $p_t$ spectra ($n_{ch}$ dependence), angular correlations ($p_t$ dependence) and minimum-bias event distributions on $n_{ch}$ [based on inferred event-wise jet number $n_j(n_{ch})$]. From those elements we construct particle-number densities on the space $(p_t,n_{ch})$ (particle transverse momentum and event multiplicity) that form the basis for studying relations among a $p_{t,trig}$ condition, particle production mechanisms and event types. For a given $p_{t,trig}$ value what are the corresponding fractions of particles from soft and hard production mechanisms, and the fractions of soft and hard event types? How are those trends related to the systematics of transverse multiplicity $N_\perp$? We also consider the consequences of a complementary event-multiplicity $n_{ch}$ condition on spectra and correlations in the context of a Glauber model for pQCD applied to \pp collisions.

\section{The two-component model} \label{twocomp}

The two-component model of \pp $p_t$ spectra was initially derived from the multiplicity dependence of spectra~\cite{ppprd}. Its physical interpretation is based on a simple model of hadron production: longitudinal projectile-nucleon dissociation (soft) and transverse large-angle-scattered parton fragmentation (hard). In more-peripheral \aa collisions hadron production from the two processes scales as $N_{part}/2$ (number of  nucleon participant pairs) and $N_{bin}$ (number of binary \nn collisions) respectively~\cite{kn,anomalous}. Those parameters are related to observable $n_{ch}$ via the Glauber model of \aa geometry~\cite{centrality}. For \pp collisions the corresponding geometry parameters and their relation to observables are considered in Sec.~\ref{gglauber}.
Interpretation of model elements in terms of physical mechanisms is fairly reliable for \pp collisions but is less certain for more-central \auau collisions. 
The TCM can also be applied to single-particle yields  and two-particle correlations. We require that all TCM interpretations be consistent across spectra, correlations, minimum-bias distributions on $n_{ch}$ and particle yields.


\subsection{p-p $\bf p_t$ spectrum systematics vs $\bf n_{ch}$}

In Ref.~\cite{ppprd} the structure of $p_t$ spectra for several charge-multiplicity $n_{ch}$ classes of \pp collision events was analyzed. The spectrum data are well-described  by a two-component model function of the form
\bea \label{pttcm}
dn_{ch}/y_tdy_t  &=&  n_s( n_{ch}) S_0(y_t)
+  n_{h}( n_{ch}) H_0(y_t)
\eea
(within rapidity acceptance $\Delta \eta = 1$), where unit-integral factors $S_0(y_t)$ and $H_0(y_t)$ (defined in App.~\ref{tcmmodels}) are respectively a L\'evy distribution on $m_t$ (suitably transformed to $y_t$) and a peaked distribution on $y_t$ approximated for this analysis by a Gaussian,. Transverse rapidity $y_t =\ln\{(m_t + p_t) / m_h\}$ is adopted to aid visualization of low-$p_t$ spectrum structure, and pion mass $m_h \rightarrow m_\pi$ is assumed for unidentified hadrons.  Soft $n_s$ and hard $n_h$ multiplicities observed within $\Delta \eta = 1$ approximate densities $dn_x/d\eta$. Quantities $H_{pp}(y_t) = n_h H_0(y_t)$ and $S_{pp}(y_t) = n_s S_0(y_t)$ are also defined. Equation~(\ref{pttcm}) represents approximate factorization of the multiplicity and $y_t$ dependence, effectively in the form of a two-term Taylor expansion on $n_{ch}$. Series coefficients for higher terms as determined by spectrum data are consistent with zero.

 \begin{figure}[h]-
  \includegraphics[width=1.62in,height=1.6in]{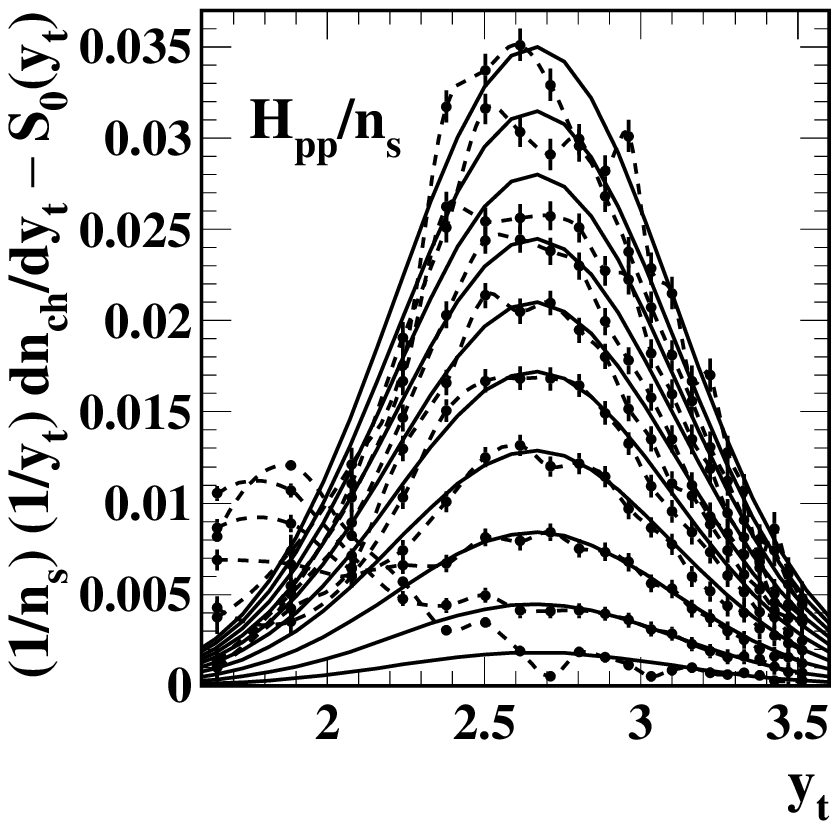}
  \includegraphics[width=1.62in,height=1.6in]{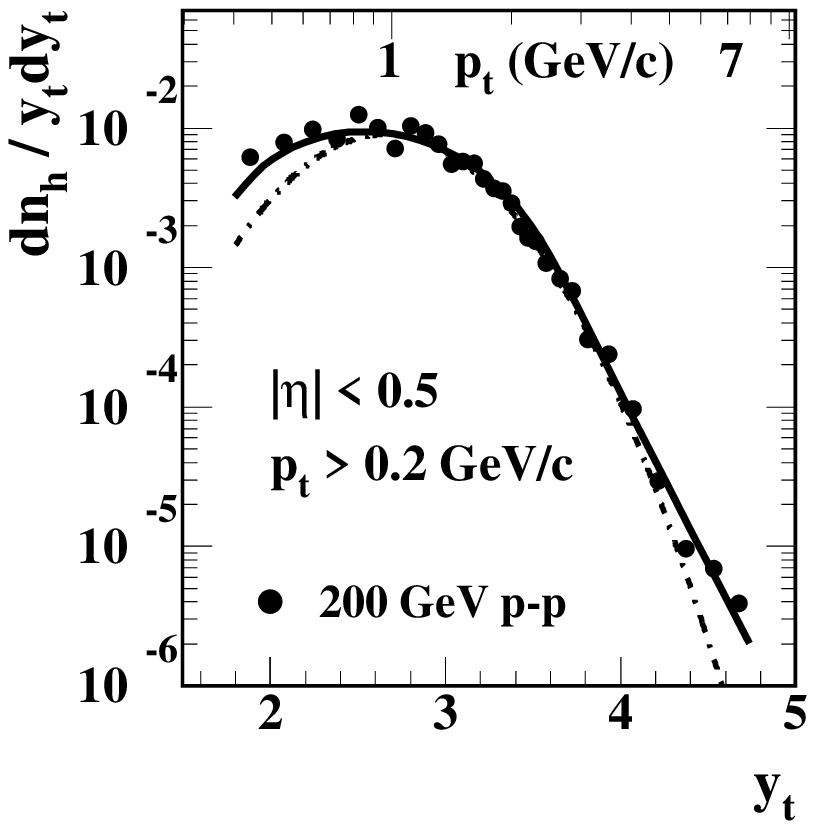}
\caption{\label{ppspec}
Left: Evolution of the $p_t$ spectrum hard component with event multiplicity $n_{ch}$, from 200 GeV \pp collisions~\cite{ppprd}.
Right:  Calculated pQCD fragment distribution (solid curve) corresponding to the NSD-equivalent hard component from the left panel (points)~\cite{fragevo}. The dash-dotted curve is the corresponding Gaussian form from the left panel (solid curves).
}  
 \end{figure}

Figure~\ref{ppspec} (left panel) shows the most important result from Ref.~\cite{ppprd}. Data in the left panel were obtained by dividing each $p_t$ spectrum by the soft-component yield $n_s(n_{ch})$ (inferred iteratively) and subtracting the fixed soft-component function $S_0(y_t)$ derived in the limit $n_{ch} \rightarrow 0$. The result is inferred hard components $H_{pp}/n_s$.  The data are consistent with $n_h / n_s \propto n_{ch}$ and $n_h \ll n_s$.

Figure~\ref{ppspec} (right panel) shows a combination of data from the left panel corresponding to non-single-diffractive (NSD) \pp collisions (points). The solid curve through the data is derived by combining published fragmentation functions~\cite{eeprd} with a pQCD parton energy spectrum~\cite{fragevo}. The dash-dotted curve is the simplified Gaussian form from the left panel, defined in App.~\ref{tcmmodels} and used in this analysis.  The good agreement between pQCD theory and \pp spectrum data strongly supports interpretation of the spectrum hard component as a jet {\em fragment distribution} representing a minimum-bias parton energy spectrum bounded below at 3 GeV. We conclude from Ref.~\cite{ppprd} that  varying the \pp $n_{ch}$ condition controls the jet frequency. Jet number $n_j$ is proportional to $n_s n_{ch} \approx n_{ch}^2$, and the constant of proportionality is consistent with pQCD and angular acceptance $\Delta \eta = 1$.


\subsection{p-p two-particle correlations}

Extensive measurements of \pp two-particle angular correlations on $(\eta,\phi)$ and correlations on transverse rapidity space $(y_{t1},y_{t2})$ for several charge combinations have been reported~\cite{porter2,porter3}. Most of the resulting structure is well-described by the TCM. The exceptions are contributions from Bose-Einstein correlations and electron pairs from photon conversions. The soft component appears mainly below $p_t = 0.5$ GeV/c whereas the hard component appears only above 0.35 GeV/c. Thus, a cut at 0.5 GeV/c imposed on both particles of each pair effectively separates hard and soft components. The hard component on $(y_{t1},y_{t2})$ projected onto 1D $y_t$ is compatible with the structure in Fig.~\ref{ppspec}, and the corresponding shapes on $(\eta,\phi)$ are as expected for interjet and intrajet angular correlations, strongly supporting a jet interpretation for the correlation hard component. \pp angular correlation data are discussed further in Sec.~\ref{angcorr}.

\subsection{Two-component model and \aa collisions}

The present study emphasizes jet production and the role of collision centrality in \pp collisions. Centrality trends in \aa collisions have been studied extensively. The relevant centrality observable for  \aa collisions is $n_{ch}$, and the corresponding Glauber centrality measure is mean participant path length $\nu \equiv 2N_{bin} / N_{part}$. The two parameters are related by the total cross section~\cite{centrality}.  By analogy with \aa collisions $n_{ch}$ may also be the preferred control parameter in \pp collisions {\em if}\, centrality is a relevant concept there.
Whereas dijet production is observed to scale as $n_{ch}^2$ in \pp collisions it scales as $N_{bin}$ in more-peripheral \aa collisions (as expected for linear superposition of \nn collisions). From the Glauber model we obtain $N_{bin} \approx N_{part}^{4/3}$ and $N_{part} \propto n_{ch}$ (approximately). Therefore dijet production scales as $n_{ch}^{4/3}$ in more-peripheral \aa collisions, not $n_{ch}^2$ as in \pp collisions. Scaling issues are discussed further in Sec.~\ref{pppcent}.

\subsection{Two-component theory description}

Reference~\cite{ppcent2} introduces an alternative two-component model for transverse multiplicity $N_\perp$ systematics based on a two-state system of large-$b$ events with no jets (soft) and small-$b$ events with jets (hard). The form is
\bea \label{nperptheory}
N_\perp(p_{t,trig}) \hspace{-.02in} &=& \hspace{-.02in} \lambda_s(p_{t,trig}) N_s(b_s) +  \lambda_h(p_{t,trig}) N_h(b_h)
\eea
with fractions $\lambda _h + \lambda_s = 1$ of hard and soft events respectively for a given $p_{t,trig}$ and fixed numbers $N_s$ and $N_h$, with $N_h \gg N_s$ assumed. While the form of that equation is similar to Eq.~(\ref{pttcm}) the underlying assumptions are quite different. By definition of $N_\perp(p_{t,trig})$ (complementary to the triggered dijet) both $N_s$ (larger $b$) and $N_h$ (smaller $b$) include {\em soft} (nonjet) components from the two event types. An MPI contribution to $N_h$ is also possible. The subscripts refer to the event types with their different centralities, not soft and hard components of a given event.  Fractional abundances $\lambda_x$ of the two event types are assumed to be determined only by $p_{t,trig}$. 

In the present analysis we define two types of \pp collision: soft and hard events. Equation~(\ref{pttcm}) applies to any \pp event. Hard events contain at least one minimum-bias jet within some specified detector angular acceptance. Soft events contain no jet structure within the acceptance ($n_h \approx 0$). All events include a common nonjet soft component of particle production. The relation of soft and hard events to \pp collision centrality (if any) is not assumed a priori. We compare theory assumptions with data trends and find that the soft components of $N_\perp$ for the two event types are actually similar, but $N_\perp$ for hard events includes an additional {\em triggered}-jet-related (hard) component that may even dominate $N_\perp$.

\section{Transverse multiplicity $\bf N_\perp(p_t)$} \label{cdftrans}

We now apply the two-component model to study transverse multiplicity $N_\perp(p_t)$ spectrum structure and other systematic issues. We assume that measured projectile dissociation (beam-beam remnants) contributes to the $N_\perp$ spectrum according to the TCM but also consider a possible hard-component contribution that may arise from the triggered dijet. 
We find that the TCM provides an accurate description of $N_\perp(p_t)$ spectra and leads to an alternative interpretation of  UE composition.

\subsection{$\bf N_{\perp}(p_t)$ spectrum structure}

\begin{figure}[h]
  \includegraphics[width=3.3in,height=1.6in]{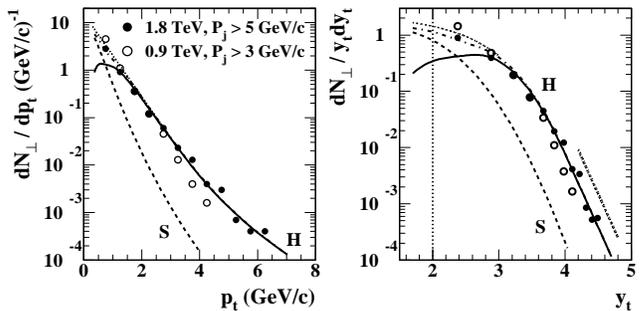}
\caption{\label{spectrum}
Left: 
Transverse multiplicity $N_\perp$ spectrum on $p_t$ for 1.8 TeV \pp collisions (solid points, $|\eta|< 1$) from Fig.~37 of Ref.~\cite{under5} and for 0.9 TeV \pp collisions (open points, $|\eta|< 2$) from Fig.~4 of Ref.~\cite{under}. Curve $S$ is the soft component with data normalization corresponding to a soft-component density $dn_{s}/d\eta = 5$.  Curve $H$ is derived from the pQCD-calculated fragment distribution for minimum-bias jets from 200 GeV \pp collisions~\cite{fragevo} also normalized to match the 1.8 TeV data.
%
Right: Data and curves from the left panel transformed to transverse rapidity $y_t$ with the appropriate Jacobian. A factor $1/y_t$ is added to compare with Fig.~\ref{ppspec} (right) and Eq.~(\ref{pttcm}).
}  
 \end{figure}

Figure~\ref{spectrum} shows transverse multiplicity $N_\perp(p_t)$  data for 1.8 TeV (solid points) for a specific trigger condition (summed jet momentum $P_j > 5$ GeV/c) from Fig.~37 of Ref.~\cite{under5}. The data can be described with TCM Eq.~(\ref{pttcm}) modulo transformation $y_t \rightarrow p_t$ and additional factor $p_t$.  The curves are unit-normal forms $p_t\, S_0(p_t)$ and $p_t\, H_0(p_t)$ derived in Ref.~\cite{ppprd} from $\sqrt{s} = 200$ GeV \pp data. The respective coefficients are $N_{\perp,s} = 3.3 \approx 2\times 5/3$ (dashed curve) and $N_{\perp,h} = 1.3 \approx 4/3$ (solid curve). The sum (dash-dotted curve) describes the 1.8 TeV data well. Data for 0.9 TeV with the lower 3 GeV/c trigger condition (open points) are discussed in Sec.~\ref{lowspec}.

The Eq.~(\ref{pttcm}) coefficients (with systematic uncertainties) were determined from the data as follows. The spectrum shapes $S_0(p_t)$ and $H_0(p_t)$ inferred from the 200 GeV data in Ref.~\cite{ppprd} (see App.~\ref{tcmmodels}) were left unchanged. Coefficient $N_{\perp,h} \rightarrow 1.3$ was adjusted to fit the CDF data (solid points) at larger $p_t$, implying that $n_{h} = 3 \times 1.3 \approx 4 \pm 0.5$ is the hard-component fragment integral within $2\pi$ azimuth and $\Delta \eta = 2$, interpreted as the minimum jet fragment multiplicity for imposed conditions (e.g., at least one jet). Coefficient $N_{\perp,s} = 3.3 \approx 2\times 5/3$ was then adjusted to fit the data at smaller $p_t$ dominated by the lowest point. The corresponding nonjet soft density $dn_{s}/d\eta \approx 5 \pm 1$ can be compared with the NSD soft multiplicity density $dn_{s}/d\eta \approx 4$ measured at 1.8 TeV~\cite{under5}.

Figure~\ref{spectrum} (right panel) shows the same data and curves on transverse rapidity $y_t$ with factor $1/y_t$ added. While $p_t$ may be a directly measured quantity it obscures details at smaller $p_t$ as compared with $y_t$. The linear power-law trend evident at larger $y_t$ (compare with the dotted line at right) is consistent with an underlying minimum-bias pQCD parton spectrum~\cite{fragevo}. 

The dash-dotted curve in the right panel can be compared with Fig.~10 (right panel) of Ref.~\cite{ppprd}  predicting the TCM spectrum shape for 200 GeV \pp hard events, those with at least one jet (dotted curve). The soft component at 1.8 TeV is increased relative to 200 GeV by 5/2.5 = 2 times whereas the minimum-bias hard component (jets) may be more slowly varying with collision energy. Although the minimum in the hard-event spectrum from Ref.~\cite{ppprd} does not appear in Fig.~\ref{spectrum} (right panel) the two results are mutually consistent within uncertainties.

\subsection{Collision energy comparisons}

In Fig.~\ref{spectrum} (right panel) the vertical scale range is shifted by a factor 30 higher than that for Fig.~\ref{ppspec} (right panel) to explore the relation between the two data sets. The jet frequency inferred for the untriggered 200 GeV data in Fig.~\ref{ppspec}  is $f = 0.013$~\cite{ppprd}, and the spectrum is integrated over $2\pi$ azimuth and $\Delta \eta = 1$. The estimated most-probable jet energy is 3 GeV. The triggered 1.8 TeV data in Fig.~\ref{spectrum} with trigger-jet momentum $P_j > 5$ GeV/c correspond to jet frequency $f \approx 1$ and are integrated over $2\pi / 3$ azimuth and $\Delta \eta = 2$. Setting aside the difference in  $\Delta \eta$ the nominal relation between data sets is then $1/(3 \times 0.013)  \approx 25$~\cite{jetspec}. The functional form derived from pQCD (solid curve,~\cite{fragevo}) appears to describe the TR hard-component spectrum accurately. 

The hard-component maximum value in Fig.~\ref{ppspec} (right panel) is about 0.01 whereas in Fig.~\ref{spectrum} (right panel) it is about 0.4, with identical shapes. The ratio 40 should be compared with the factor 25 derived above.  The hard-component yield at 200 GeV is {\em apparently} about 60\% of that at 1.8 TeV. But there are two related issues: (a) effects of different $\eta$ acceptances and (b) bias effects from a 5 GeV/c jet trigger vs no trigger. The apparent factor-1.6 difference may simply arise from differences in the jet $\eta$ acceptance and effective mean jet energy. Those issues are discussed further in Secs.~\ref{minbiasjet} and \ref{transverse}.

\subsection{Spectrum extrapolation}

$N_\perp(p_t)$ spectrum extrapolation in Sec.~V of Ref.~\cite{under5} assumes a monolithic spectrum approximated by $\exp(-2 p_t)$ (dotted curves in Fig.~\ref{spectrum}).  For plateau value $N_\perp = 2.3$ the inferred $\eta$ density is $dn_{ch}/d\eta = 3.5$ with $p_t > 0.5$ GeV/c. Extrapolation to $p_t = 0$ with the assumed spectrum shape gives $dn_{ch}/d\eta \approx 10$ compared with soft component $dn_{s}/d\eta \approx 4$. The ratio 2.5 suggests a substantial unexplained contribution to the UE~\cite{under5}.

However, accurate extrapolation of any $p_t$ spectrum  requires knowledge of its structure. Measured \pp spectra at 200 GeV can be decomposed into soft and hard components with known spectrum shapes, as in App.~\ref{tcmmodels}. We apply the same TCM to $N_\perp(p_t)$ assuming the component shapes are similar at 1.8 TeV. As noted, the soft-component coefficient derived from the spectrum data in Fig.~\ref{spectrum} corresponds to $dn_{s}/d\eta \approx 10/2 = 5$.  The hard-component coefficient $N_{\perp,h}$ corresponds to $dn_h/d\eta \approx 4/2 = 2$. The integral of the apparently unexplained (hard) part of the UE is thus only 2/7 $\approx$ 30\% of the total, and according to the TCM that component may be identified in some sense with jet production.

The most surprising aspect of the UE $p_t$ spectrum is not a factor-2.5 ratio of extrapolated $N_\perp$ to that expected for soft events. It is the {\em factor-20} ratio at $y_t \approx 3.3$ ($p_t \approx 2$ GeV/c) between the $N_\perp(p_t)$ spectrum (solid curves) and the {\em expected} soft spectrum component (dashed curves) in Fig.~\ref{spectrum}. That result strongly suggests that a large jet-related (hard) contribution may dominate the UE. The origin of the hard contribution to $N_\perp$ (MPI or triggered dijet) is a central issue for this study.

\section{$\bf p$-$\bf p$ minimum-bias distribution} \label{minbias}


Equation~(\ref{pttcm}) describes the $p_t$ spectrum structure for a given event multiplicity $n_{ch}$. We also require a model that describes the minimum-bias event distribution on $n_{ch}$ for 200 GeV \pp collisions and its soft and hard components. We adopt a model consistent with measured energy systematics up to 7 TeV. We then use the observed jet production rate vs $n_{ch}$ from Ref.~\cite{ppprd} to determine the fraction of soft events $P_0(n_{ch})$ (no jet in the acceptance) and of hard events $1 - P_0(n_{ch})$ (at least one jet in the acceptance), with $P_0(n_{ch}) = \exp[-n_j(n_{ch})]$ and $n_j(n_{ch})$ the mean jet number in angular acceptance $\Delta \eta$ for event multiplicity $n_{ch}$ as reported in Sec.~\ref{nj}.
Quantity $N_t = \sigma_X {\cal L}t$ represents the total number of events in an ensemble for some trigger condition X (e.g., NSD) and integrated luminosity ${\cal L}t$. The unit-normal event distribution on $n_{ch}$ is represented by $(1/N_t) dN_t/dn_{ch}$.

 \begin{figure}[h]
  \includegraphics[width=1.65in,height=1.6in]{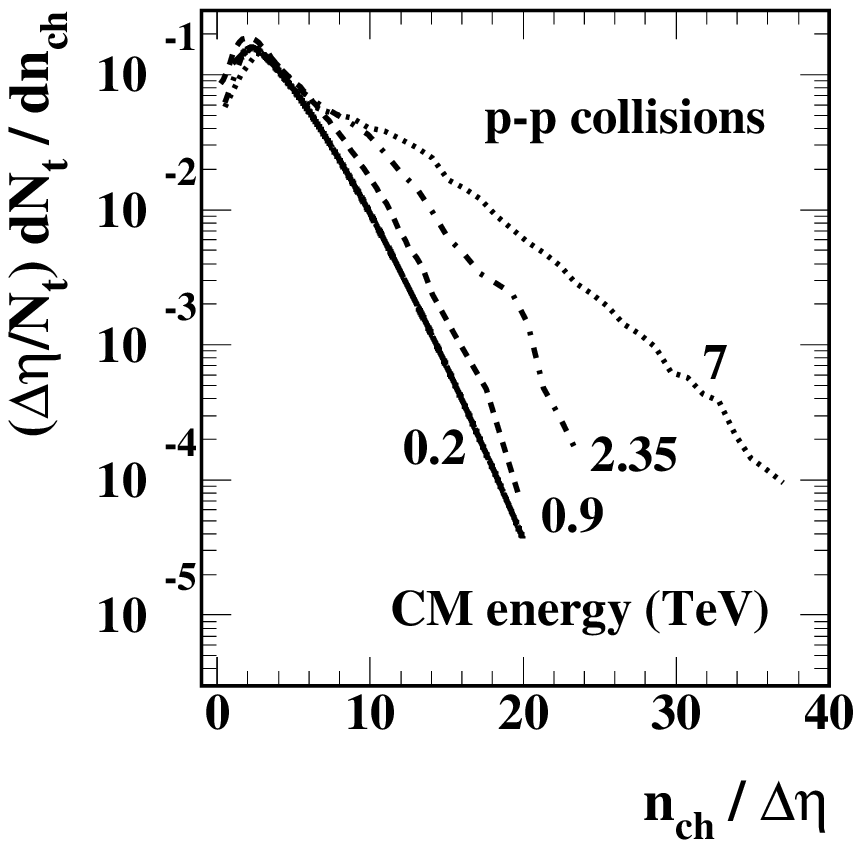}
  \includegraphics[width=1.65in,height=1.6in]{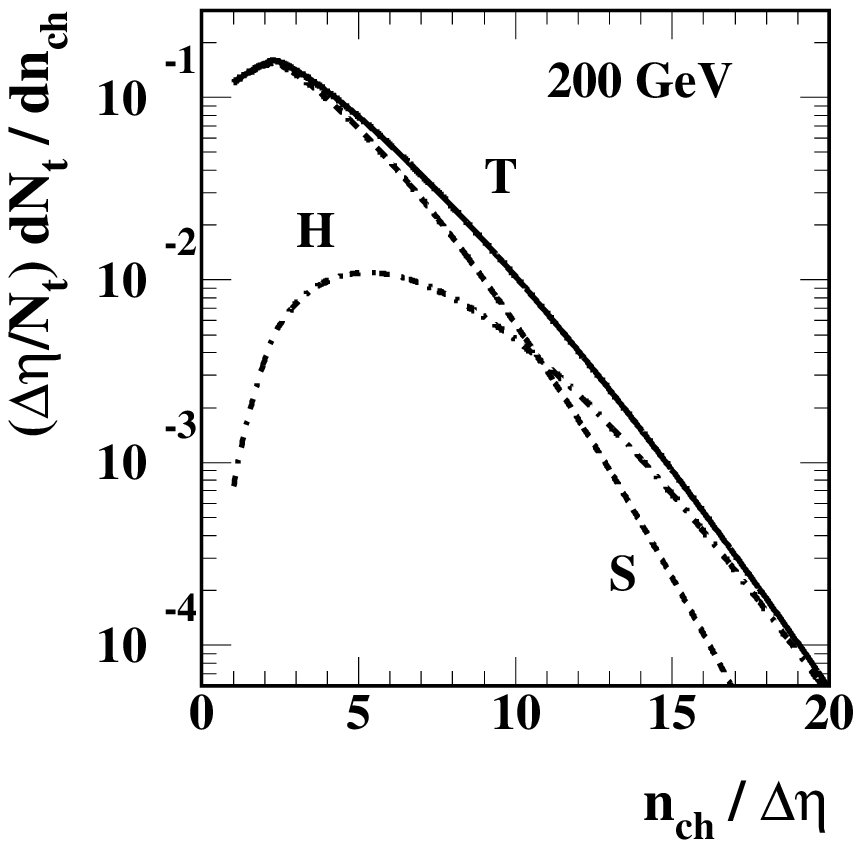}
\caption{\label{ppmult}
Left: Measured minimum-bias distributions on $n_{ch}$ for three energies (broken curves) from Ref.~\cite{cmsmult} with $|\eta| < 2.4$ and an extrapolation to 200 GeV (solid curve).
Right: The 200 GeV extrapolation $T$ is separated into soft $S$ and hard $H$ components based on results from Ref.~\cite{ppprd} (event-wise jet frequency $f$) providing a two-component model of the minimum-bias distribution.
}  
 \end{figure}

Figure~\ref{ppmult} (left panel) shows distributions on $n_{ch} / \Delta \eta$ from the CMS collaboration (broken curves) for three collision energies~\cite{cmsmult}. The data approach asymptotically the solid curve estimating the minimum-bias distribution for 200 GeV collisions. The 200 GeV parametrization is
\bea
\frac{\Delta \eta}{N_t}\frac{ dN_t} {dn_{ch}} = 0.16 \exp(-|n_{ch}/\Delta \eta - 2.3|^{1.3}/5).
\eea
Although an approximation to the data, the model function is convenient and suffices for the present study.
Figure~\ref{ppmult} (right panel) shows an expanded view of the 200 GeV model with the soft and hard components. $N_s$ is the number of soft events (distribution S), and $N_h$ the number of hard events (distribution H) with $N_t = N_s + N_h$ based on $P_0(n_{ch})$ defined above. 


The format of Fig.~\ref{ppmult} with accepted $n_{ch}$ normalized by $\Delta \eta$ contrasts with that in Ref.~\cite{cmsmult} where {total} multiplicities from different $\eta$ acceptance intervals are compared. Use of $n_{ch} / \Delta \eta$ (consistently on both abscissa and ordinate) may aid comparison of data from different acceptances and energies. 
A range of $\eta$ acceptances (e.g., $\Delta \eta = 1 \rightarrow 4.8$ in Ref.~\cite{cmsmult}) implies nearly a factor-two variation in the number of jets per detected dijet (i.e., the probability that a jet within the $\eta$ acceptance also has the recoil partner within the acceptance)~\cite{jetspec}.
Figure~\ref{ppmult} can be compared with the so-called KNO format~\cite{kno} wherein $\bar n P_n$ is plotted vs $n / \bar n$ [$P_n$ is here $(1/N_t) dN_t/dn_{ch}$]. The acceptance $\Delta \eta$ does cancel in both product and ratio, but the mean $\bar n_{ch}$ may include contributions from two or more production mechanisms, making overall interpretation difficult.



\section{Total particle density on ($\bf y_t,n_{ch}$)} \label{ttotal}

We have defined two-component models for particle multiplicity distributed on $y_t$ as spectra $dn_{ch} /y_t dy_t$, and for event number distributed on event multiplicity as $dN_t/dn_{ch}$. We now combine those results to produce a 2D distribution of total particle multiplicity $N_{ch} = \bar n_{ch} N_t$ (defining ensemble-mean event multiplicity $\bar n_{ch}$) on the space $(y_t,n_{ch})$. The 2D particle distribution can be decomposed into soft and hard components in two ways: according to hadron production mechanism (particle origin) and according to event type.  A distinction should be maintained between soft and hard particle production processes and soft and hard event types. A hadron from a soft process may emerge from a hard event, but the reverse is not possible.

In the following we assume that a trigger particle for given $y_{t,trig}$ bin is represented on average by the most-probable particle in that $y_t$ bin from all events in the ensemble, which then defines the most-probable event $n_{ch}$ (the mode) for that condition. Those events are most likely to produce the required trigger particle.  

\subsection{Description according to particle type}


Decomposition according to {particle type}  (hadron production mechanism) can be expressed by
\bea \label{2ddens}
\frac{dN_{t}}{dn_{ch}}\frac{dn_{ch}}{y_t dy_t}
&=& \frac{dN_{t}}{dn_{ch}} n_s(n_{ch}) S_0(y_t)   \nonumber \\
&+& \frac{dN_{t}}{dn_{ch}} n_h(n_{ch}) H_0(y_t),
\eea
where a common factor $\Delta \eta/N_t$ is suppressed, $n_s = n_{ch} - n_h$ and $n_h = n_j(n_{ch})\, \bar n_{ch,j}$ is the hard component of event multiplicity (jet fragments). $dn_{ch}$ appearing in a denominator and referring to an event index should be distinguished from $dn_{ch}$ in a numerator denoting a differential element of event multiplicity. 

\begin{figure}[h]
  \includegraphics[width=1.65in,height=1.6in]{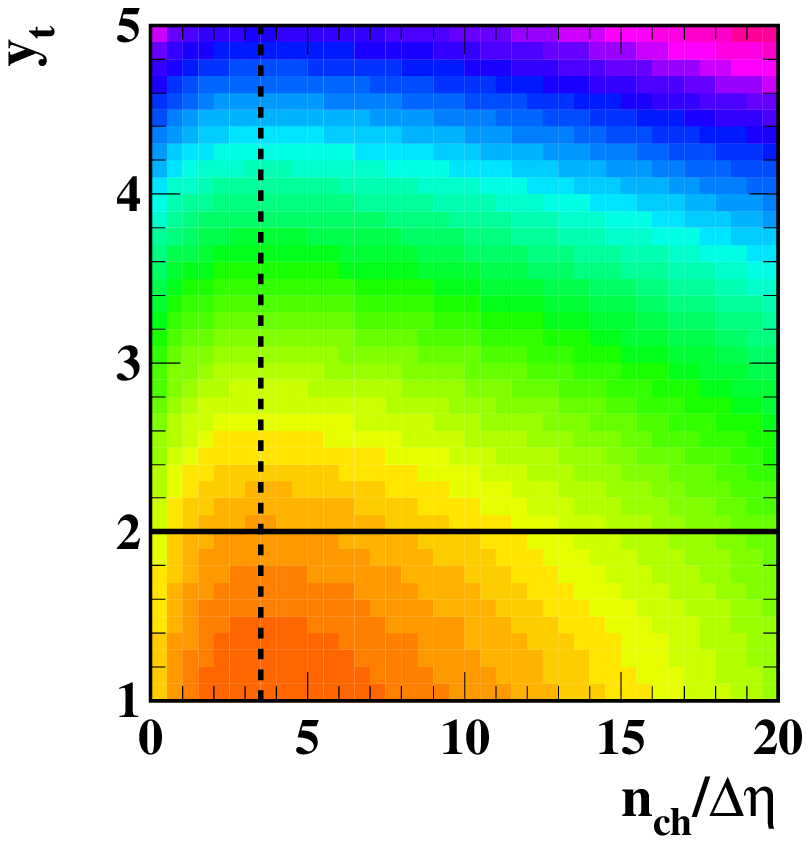}
  \includegraphics[width=1.65in,height=1.6in]{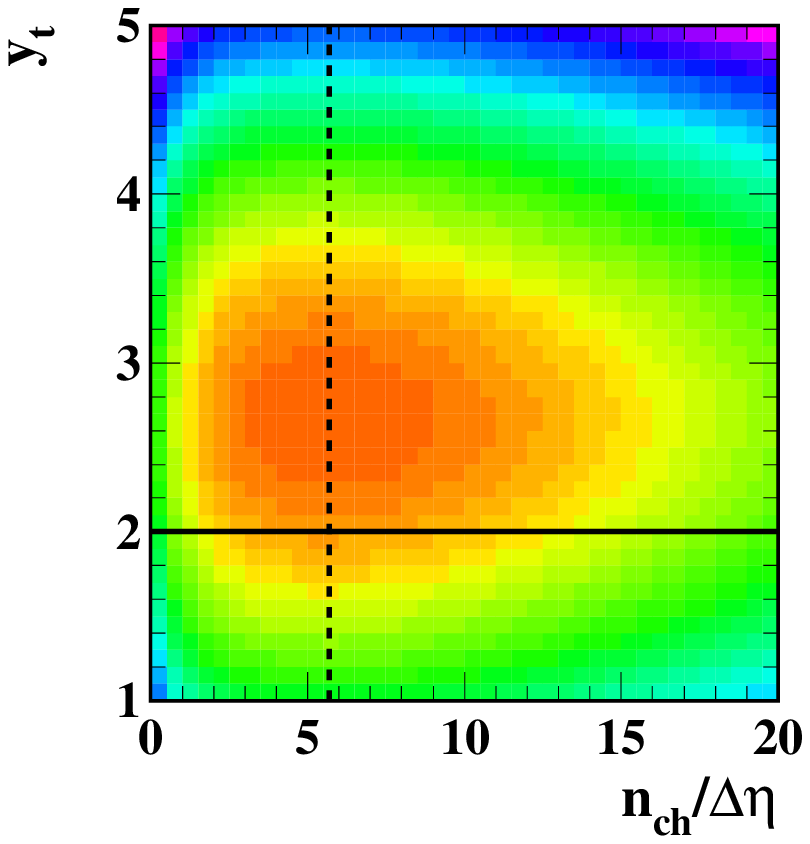}
\caption{\label{cartprod}
(Color online) Left: 2D surface representing the Cartesian product of two 1D distributions, the first term (soft) of Eq.~(\ref{2ddens}).
Right:  Surface representing the second term (hard) of Eq.~(\ref{2ddens}). The dashed lines represent the modes on $n_{ch} / \Delta \eta$ at 3.5 and 5.7.
}  
 \end{figure}

Figure~\ref{cartprod} shows surface plots of the soft component (left panel) and hard component (right panel) of the total multiplicity distribution on $(y_t,n_{ch})$ defined by Eq.~(\ref{2ddens}) (first and second terms respectively). The z-axis scales are logarithmic. Dashed lines indicate soft and hard distribution modes (most probable values) $n_{ch,s}/\Delta \eta = 3.5$ and  $n_{ch,h}/\Delta \eta = 5.7$. A condition imposed on $y_t$ would control the distribution shape on $n_{ch}$, whereas a condition on $n_{ch}$ would control the $y_t$ spectrum shape. We want to determine what event class (e.g., $n_{ch}$ value) is most probable for a specific trigger condition on $p_t$ or $y_t$.

\begin{figure}[h]
  \includegraphics[width=1.65in,height=1.6in]{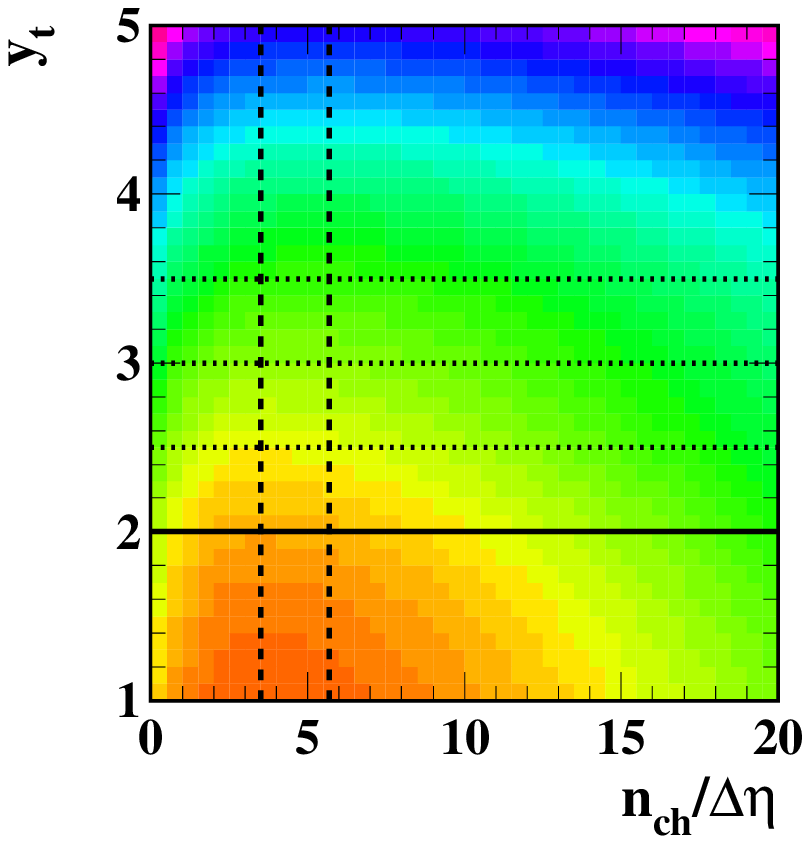}
  \includegraphics[width=1.65in,height=1.6in]{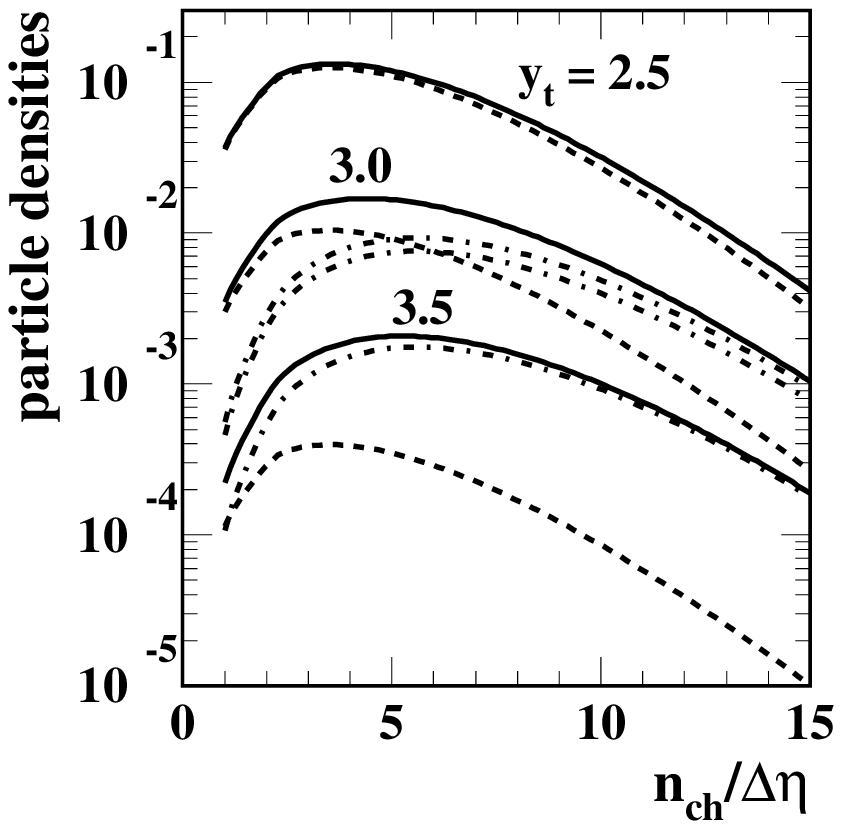}
\caption{\label{total}
(Color online) Left:  2D surface representing the sum of soft and hard terms in Eq.~(\ref{2ddens}). The dashed lines indicate the soft and hard modes on $n_{ch}/\Delta \eta$. The dotted lines represent three $y_t$ conditions for illustration.
Right: Corresponding particle densities on event multiplicity $n_{ch}/\Delta \eta$ for the  three conditions (cuts) on $y_t$ (with $p_t \approx 0.9$, 1.4 and 2.2 GeV/c respectively). Solid curves (labeled by $y_t$ cut) represent total densities. Dashed curves represent soft components and dash-dotted represent hard components.
}  
 \end{figure}

Figure~\ref{total} (left panel) shows the sum of the two terms in Eq.~(\ref{2ddens}). The vertical dashed lines indicate the modes on $n_{ch} / \Delta \eta$ of the soft and hard components. It is notable that the separation is approximately 2.5, the mean fragment multiplicity of a minimum-bias jet or minijet. 
The dotted lines indicate three $y_t$ conditions employed in discussion. The bold solid line indicates the CDF acceptance cutoff at $p_t = 0.5$ GeV/c.

Figure~\ref{total} (right panel) shows conditional particle density distributions on event $n_{ch}$ as defined by Eq.~(\ref{2ddens}) for three conditions (trigger cuts) on $y_t$. For the lowest cut the total density is dominated by the soft component (dashed). For the highest cut the total is dominated by the hard component (dash-dotted). Variation of a $p_{t,trig}$ condition over a rather small interval can change the particle-type mixture from almost all soft-component hadrons to almost all hard-component hadrons. However, the soft component of the event multiplicity (possibly related to \pp centrality) does not change significantly. That panel can be compared with the spectrum trends in Fig.~1 of Ref.~\cite{ppprd} and in Fig.~\ref{ppspec} (left panel) of this paper where $n_{ch}/\Delta \eta$ is the condition variable and the soft-component yield varies over a large range. 


\subsection{Description according to event type} \label{etype}

In Ref.~\cite{ppcent2} it is proposed to control the fraction of events that include hard processes (hard events) by specifying a $p_t$ trigger condition. With increasing $p_{t,trig}$ the fraction of hard events should increase. Assuming a strong correlation among jet production, \pp centrality and soft multiplicity there should also be a substantial increase of transverse multiplicity $N_\perp$ (assumed comparable to the multiplicity soft component of the TCM) up to some critical $p_{t,trig}$ value. Beyond that value accepted collisions should be mainly hard, nearly central and therefore with similar characteristics including $N_\perp$. 



In Eq.~(\ref{2ddens}) the 2D particle density is decomposed according to hadron production mechanism.
Decomposition according to {\em event type} can be expressed by
\bea \label{softeq}
\frac{\Delta \eta\,  d^2N_{ch,s}}{dn_{ch}y_t dy_t} &=& P_0(n_{ch}) \frac{\Delta \eta}{N_t} \frac{dN_t}{dn_{ch}} n_s(n_{ch}) S_0(y_t) \\ \nonumber
&=& C_s(n_{ch}) S_0(y_t)
\eea
for soft events (no jets), with $n_s = n_{ch}$ since $n_h = 0$, and
\bea \label{hardeq}
\frac{\Delta \eta\,  d^2N_{ch,h}}{dn_{ch}y_t dy_t} &=& [1-P_0(n_{ch})] \frac{\Delta \eta}{N_t} \frac{dN_t}{dn_{ch}} \times  \\ \nonumber
& & \left[n'_s(n_{ch}) S_0(y_t) + n'_h(n_{ch}) H_0(y_t)\right] \\ \nonumber
&=& C'_s(n_{ch}) S_0(y_t) + C_h(n_{ch}) H_0(y_t)
\eea
 for hard events (at least one jet), where $n'_s = n_{ch} - n'_h$, and $n'_h = n_h / [1-P_0(n_{ch})] \geq \bar n_{ch,j}$ is distinct from $n_h$ averaged over all soft and hard events combined. 

 \begin{figure}[h]
  \includegraphics[width=1.65in,height=1.6in]{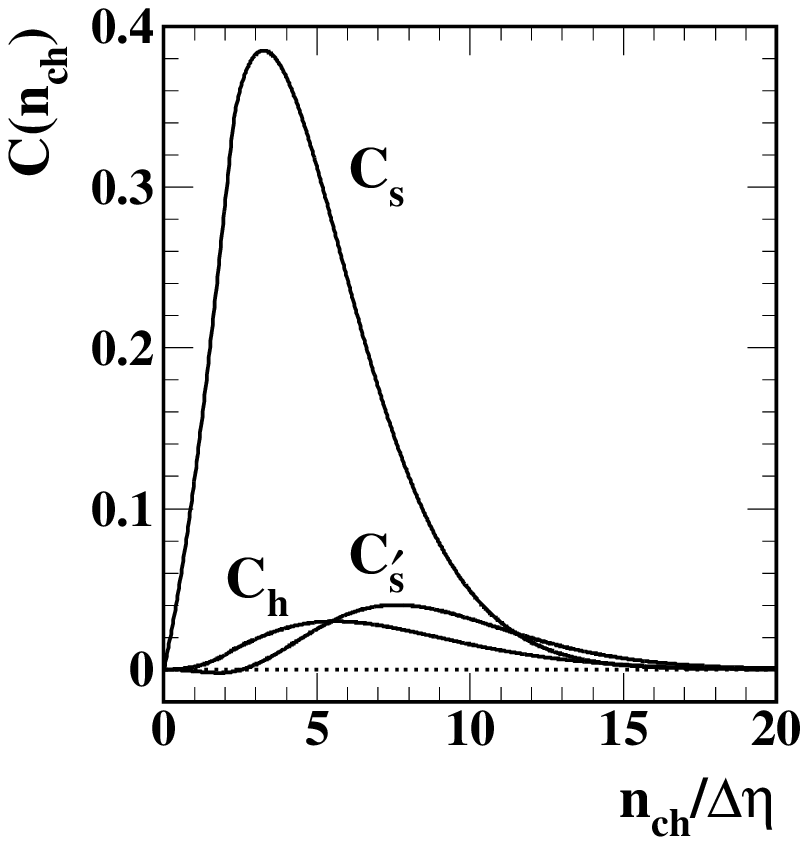}
  \includegraphics[width=1.65in,height=1.6in]{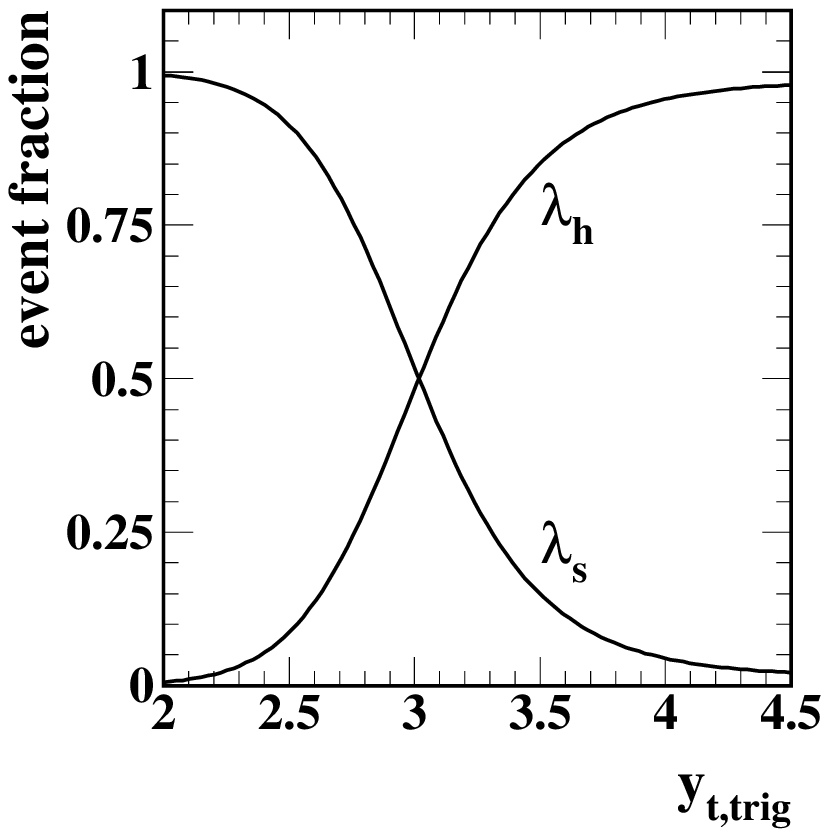}
\caption{\label{trends}
Left: Factors from Eqs.~(\ref{softeq}) and (\ref{hardeq}) that multiply $S_0$ and $H_0$ respectively. $\rho_s$ describes the soft component of soft events and $\rho'_s$ describes the soft component of hard events.
Right: 
Corresponding event-type fractions derived from Eq.~(\ref{fraceq}) corresponding to definitions in Ref.~\cite{ppcent2}.
}  
 \end{figure}

Figure~\ref{trends} (left panel) shows factors $C_s(n_{ch})$, $C'_s(n_{ch})$ and $C_h(n_{ch})$ defined by Eqs.~(\ref{softeq}) and (\ref{hardeq}). For soft events we observe that $\bar n_{ch,s} / \Delta \eta = 3.2$ and the maximum soft-event particle density $C_{s,max} = 0.39$. 
%
For hard events we obtain $\bar n_{ch,h} / \Delta \eta = 5.7$ and maximum hard-event particle density $C_{h,max} = 0.035$. 
The hard and soft event-type modes $\bar n_{ch,x}$ are formally comparable to asymptotic values $N_{soft}$ and $N_{hard}$ in Ref.~\cite{ppcent2}, but the actual hadron production mechanisms and event-type criteria may not be as assumed.


Figure~\ref{trends} (right panel) combines properties of the multiplicity densities to obtain estimated event-type fractions $\lambda_x$ vs $y_{t,trig}$ as introduced in Eq.~(\ref{nperptheory}). Based on the TCM the soft-event fraction is
\bea \label{fraceq}
\lambda_s(y_{t,trig}) \hspace{-.03in} &=& \hspace{-.03in} \frac{C_{s,max} S_0(y_{t,trig})}{C_{s,max} S_0(y_{t,trig})+ C_{h,max} H_0(y_{t,trig})}
\eea
with a corresponding expression for $\lambda_h(y_{t,trig})$ if $s,S \leftrightarrow h,H$. Those results relate to Eq.~(15) of Ref.~\cite{ppcent2}. Since $C'_s$ is at most 10\% of $C_s$ it is omitted from Eq.~(\ref{fraceq}). The event-type fractions cross at $p_{t,trig} \approx 1.5$ GeV/c, and above 2.5 GeV/c hard events comprise more than 90\% of the total. Although only 3\% of 200 GeV NSD \pp collisions are hard events (Sec.~\ref{nj}) a $p_{t,trig} > 2.5$ GeV/c condition selects almost all such events and almost no soft events.  The soft component of the hard-event ensemble is approximately that of the minimum-bias sample.

\section{Angular correlation structure} \label{angcorr}

We can use the structure of measured minimum-bias angular correlations from 200 GeV \pp collisions to predict a possible dijet contribution to the transverse azimuth region at higher collision energies, assuming that {\em minimum-bias} jets are approximately universal over a broad range of \pp collision energies.

\subsection{Minijets from $\bf p$-$\bf p$ collisions at 200 GeV}

\begin{figure}[h]
  \includegraphics[width=3.3in,height=1.6in]{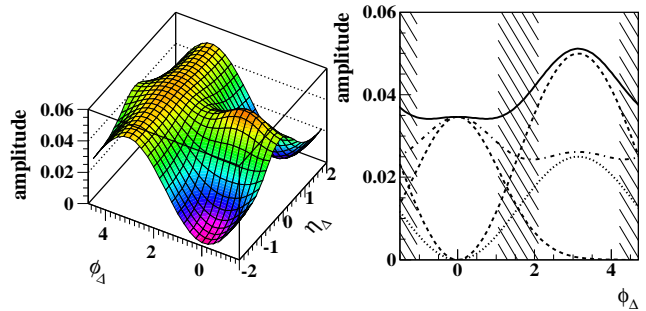}
\caption{\label{minijet}
(Color online)
Left: 
Minimum-bias jet-related 2D angular correlations from 200 GeV \pp collisions~\cite{porter2,porter3}. A same-side 2D peak at the origin is elongated on azimuth. The away-side 1D peak is broad on azimuth and approximated by a dipole $\cos{\phi_\Delta}$ form.
Right: 
Projection of jet-related angular correlations onto azimuth. The hatched regions represent the periodic ``tranverse region'' (TR) invoked in underlying-event studies. The unhatched regions, denoted ``toward'' ($\phi_\Delta \approx 0$) and ``away'' ($\phi_\Delta \approx \pi$), are typically assumed to contain all {\em triggered} dijet structure.
}  
 \end{figure}

Figure~\ref{minijet} (left panel) shows a parametrization of measured 2D angular correlations representing minimum-bias jets (minijets) in 200 GeV NSD \pp collisions from Refs.~\cite{porter2,porter3}. No trigger condition is imposed---the distribution represents all combinatoric pairs above 0.5 GeV/c (after subtraction of a uniform background based on 2D model fits).
Figure~\ref{minijet} (right panel) shows the 2D histogram projected onto 1D azimuth (solid curve).  Note that the away-side pair structure (dashed curve) has been modified by a triangular $\eta$-acceptance correction that overestimates the accepted AS pairs. The dash-dotted sum represents the uncorrected (accepted) pair density.

Both the same-side 2D peak and the away-side 1D peak are broad on azimuth. For minimum-bias jets the initial-state parton $k_t \approx 1$ GeV/c is comparable to the minimum-bias mean parton $p_t$ $\approx 3$ GeV/c. The AS azimuth r.m.s.\ width is then approximately $\pi/2$, and the periodic  AS peak array is approximated by dipole $\cos(\phi_\Delta)$ (dashed curve) extending well into the SS region~\cite{tzyam}. The SS 2D peak for minimum-bias jets in NSD \pp collisions is strongly elongated on azimuth with 2:1 aspect ratio (dash-dotted curve). Thus, the {\em untriggered} SS and AS jet peaks are {\em strongly overlapping} on azimuth.


The transverse region invoked in UE studies and indicated by the hatched regions in the right panel is conventionally assumed to contain no contribution from a triggered high-$p_t$ (di)jet (if each jet is confined to a cone of radius $R < 1$) and should therefore be particularly sensitive to the UE complementary to the dijet. 
Figure~\ref{minijet} indicates that the TR must include a substantial part of the fragment yield from {\em minimum-bias} jets. The TR is not qualitatively distinguished from any other part of the minimum-bias jet structure and may be dominated by the hard component (minimum-bias jet fragments). Since a hard event includes by definition at least one jet with mean fragment multiplicity $\bar n_{ch,j} \geq 2.5$ the TR in hard events should include a jet fragment density $d^2n_h/d\eta d\phi$ corresponding to at least that  multiplicity. 

\subsection{Minimum-bias jets at 200 GeV and 1.8 TeV} \label{minbiasjet}

\begin{figure}[h]
  \includegraphics[width=3.3in,height=1.6in]{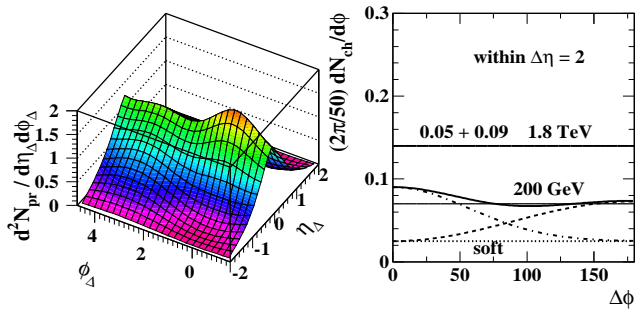}
\caption{\label{cdf}
(Color online)
Left: The histogram in Fig.~\ref{minijet} (left panel) is renormalized by the calculated jet frequency within $\Delta \eta = 2$ from 200 GeV NSD \pp collisions. The pair $\eta$-acceptance correction has also been reversed.
Right: Projection of the histogram in the left panel onto azimuth (solid and broken curves) in the format of Fig.~17 of Ref.~\cite{under5} compared to results from the CDF study (bold horizontal line).
}  
 \end{figure}

Figure~\ref{cdf} (left panel) shows the surface in Fig.~\ref{minijet} (left panel) scaled up by factor 1/0.03 $[n_j(n_{ch} =  2.5) \approx 0.03$ is the mean number of dijets in NSD \pp events within $\Delta \eta = 2$]. From spectrum analysis and comparisons with pQCD calculations the most-probable jet energy is 3 GeV ($Q = 6$ GeV)~\cite{ppprd,hardspec,fragevo}.  As noted, the triangular pair $\eta$-acceptance correction has also been removed, restoring the accepted jet-related two-particle angular correlations in hard \pp collisions, those with a triggered dijet.

Figure~\ref{cdf} (right panel) has a format similar to Fig.~17 of Ref.~\cite{under5}. The curves for 200 GeV are for minimum-bias jets (within $\Delta \eta = 1$) with 1.3 (SS) and 1.2 (AS) mean fragment multiplicities summing to 2.5, consistent with the assumed $\bar n_{ch,j}$ in Ref.~\cite{ppprd}. The lower solid line assumes $N_{\perp,h} = 2.25 / 3$. The soft background is $2 \times 2.5 \times 0.25 / 50 = 0.025$. 
Those numbers can be compared with triggered 5 GeV jets at 1.8 TeV where the computed soft background is $2 \times 5 \times 0.25 /50 = 0.05$ and the UE hard component is $ (6.9 - 10\times 0.25) /  50 \approx 0.09$. The hard component inferred from the $N_\perp(p_{t,trig})$ plateau value 2.3 is thus $n_h = 4.4$, whereas that inferred from the $N_\perp$ spectrum in Fig.~\ref{spectrum} is $n_h = 4$ (assuming the same soft component for both cases). The two numbers are consistent within systematic uncertainties.
The 200 GeV estimate only establishes a {\em lower limit}. Loss of jet fragments at the $\eta$ acceptance edges for $\Delta \eta = 1$ was not considered in the 200 GeV spectrum analysis. Such losses may be as much as 50\% for $\Delta \eta = 1$ and 25\% for $\Delta \eta = 2$. A larger fraction of partner jets is associated with triggered dijets in the larger $\eta$ acceptance. The 5 GeV/c jet trigger at 1.8 TeV may also bias jets to larger $N_\perp$ multiplicities compared to the untriggered 200 GeV study. Effects of jet bias and $\eta$ acceptance are discussed further in Sec.~\ref{transverse}.


\section{Transverse multiplicity trends} \label{transverse}

We can now predict the $N_\perp(p_{t,trig})$ or $d^2N_{\perp}/d\eta d\phi$ vs $y_{t,trig}$ trends observed in UE studies using results from 200 GeV event-type characterization and angular correlations. We combine particle density distributions for soft and hard events obtained from Eqs.~(\ref{softeq}) and (\ref{hardeq}) with soft and hard spectrum components to determine soft and hard event fractions $\lambda_x$ as functions of $y_{t,trig}$ as in Eq.~(\ref{fraceq}).  
To facilitate comparisons with previous analysis we adopt the acceptance convention $|\eta| < 1$, $p_{t}  > 0.5$ GeV/c. Acceptance fractions $\epsilon$ for soft and hard spectrum components with a given $p_{t,min}$ lower acceptance cutoff and  fractions $g$ with given $p_{t,trig}$ upper cutoff are determined below. The hard-component contribution to the TR has been estimated in the previous section, based on angular correlation measurements at 200 GeV. 
$p_{t,trig}$ (single-particle $p_t$, not jet sum $P_j$) or $y_{t,trig}$ is the trigger condition, and the total event multiplicity $n_{ch}$ in $\eta$ acceptance $\Delta \eta$ is the sum of soft $n_s$ (projectile dissociation) and hard $n_h$ (jet fragment) multiplicities.

\subsection{$\bf p_t$-cut acceptances}

\begin{figure}[h]
   \includegraphics[width=1.65in,height=1.6in]{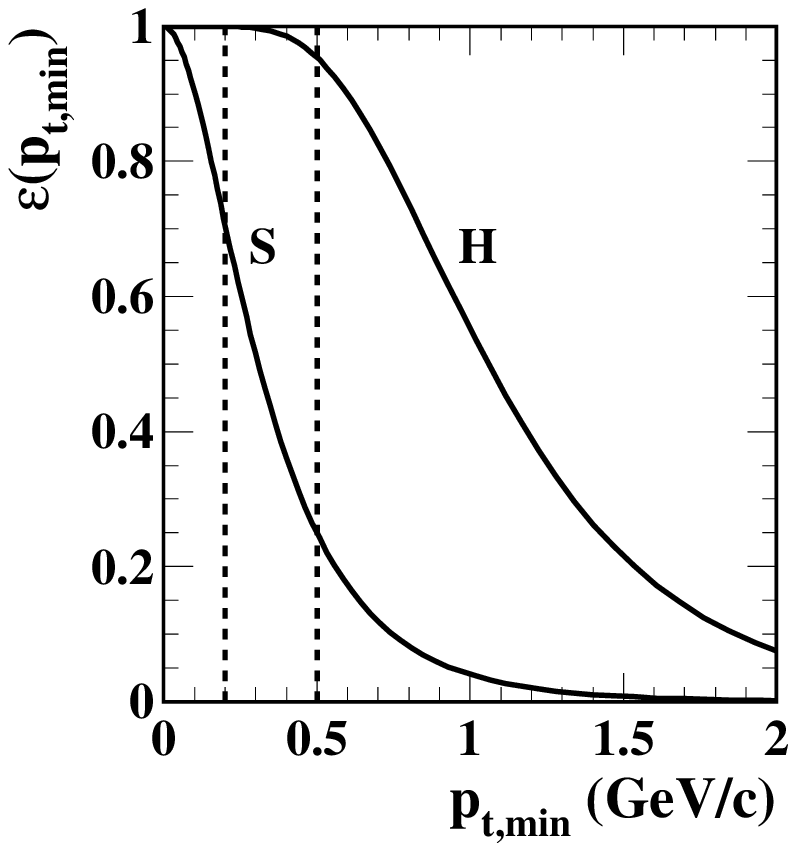}
\includegraphics[width=1.65in,height=1.6in]{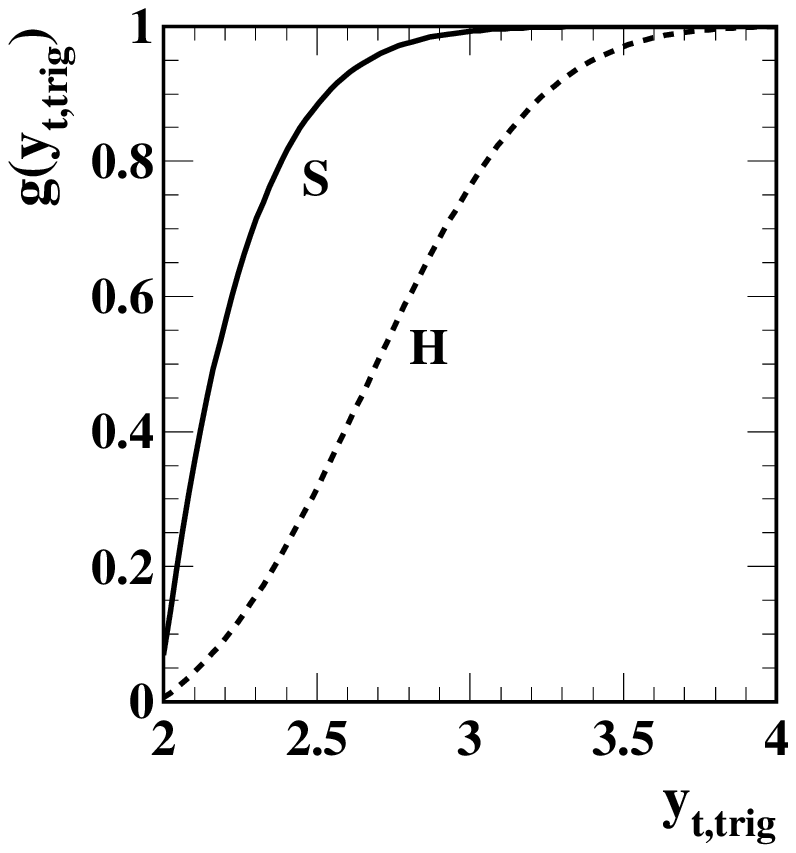}
\caption{\label{runint}
Left: 
Running integrals from below of soft $S_0$ and hard $H_0$ components defined in App.~\ref{tcmmodels}  vs $p_t$ acceptance lower cutoff $p_{t,min}$. The vertical dashed line indicates a conventional cutoff at $p_t = 0.5$ GeV/c ($y_t \approx 2$) for UE studies. 25\% of the soft component passes that cut.
Right: 
Running integrals from above of soft $S_0$ and hard $H_0$ components vs single-particle trigger condition $y_{t,trig}$ that serves as an effective upper cutoff on particle momentum.  
}  
 \end{figure}

Figure~\ref{runint} (left panel) shows the fractions $\epsilon_s$ and $\epsilon_h$ of soft and hard particle types that survive the $p_t$ acceptance lower bound $p_{t,min}$ or $y_{t,min}$. The curves are running integrals from below of functions $S_0(y_t)$ and $H_0(y_t)$ defined in App.~\ref{tcmmodels}. 
The vertical dashed lines denote the CDF acceptance boundary at 0.5 GeV/c that accepts about 25\% of the soft component and more than 95\% of the hard component, and the acceptance boundary at 0.2 GeV/c from Ref.~\cite{ppprd} that accepted 70\% of the $p_t$ spectrum (assuming 100\% tracking efficiency).

Figure~\ref{runint} (right panel) shows factors $g_s$ and $g_h$ that represent the fractions of soft and hard particles surviving the single-particle trigger condition $p_{t,trig}$ as an effective spectrum {upper bound}. 

\subsection{Transverse multiplicity systematics at 200 GeV}

Transverse multiplicity density $d^2N_\perp/d\eta d\phi$ is the mean 2D angular density within the TR. 
An expression corresponding to Eq.~(15) of Ref.~\cite{ppcent2} is
\bea \label{nperpeq}
\frac{d^2N_{\perp}}{d\eta d\phi}(y_{t,trig}) &=& \lambda_s(y_{t,trig})g_s(y_{t,trig}) \epsilon_s \bar n_{ch,s}/2\pi\Delta \eta \\ \nonumber
&+& \lambda_h(y_{t,trig})g_h(y_{t,trig}) \epsilon_h \bar n'_{ch,h}/2\pi\Delta \eta
\eea
For soft ($\approx$ NSD) events at 200 GeV and the CDF $y_t$ acceptance $\bar n_{ch,s}/\Delta \eta \approx 2.5$ and $\epsilon_s \approx 0.25$. For hard events $\bar n'_{ch,h} / \Delta \eta = 2.25/2 \approx 1.1$~\cite{ppprd}) and $\epsilon_h \approx 0.95$. For simplicity we assume that the most-probable jets have the same properties for any trigger, are not biased. With increasing $p_{t,trig}$ the triggered jet energy and fragment multiplicity may increase, but the additional higher-$p_t$ particles should appear closer to the dijet axis and therefore do not contribute significantly to $N_\perp$ within the TR. 



\begin{figure}[h]
  \includegraphics[width=3.3in,height=1.6in]{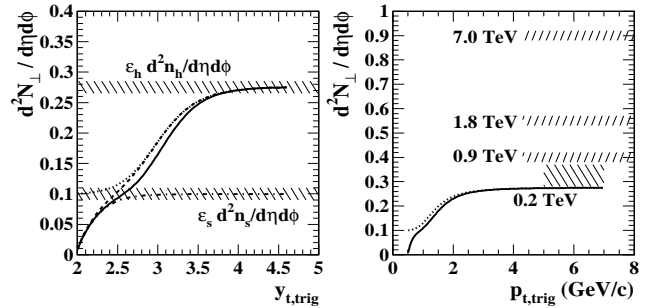}
\caption{\label{trends2}
Left: 
Transverse multiplicity angular density vs trigger condition $y_{t,trig}$ (solid curve) for 200 GeV \pp collisions. The single-particle trigger condition also serves as an upper acceptance cut on soft and hard components. The dotted curve is without that acceptance cut, the dashed curve is with the upper cut imposed on the soft component only.
Right: The same curves plotted on $p_{t,trig}$. Three labeled hatched regions summarize results from higher collision energies for comparison. The fourth hatched region estimates an uncertainty interval for the 200 GeV prediction.
}  
 \end{figure}

Figure~\ref{trends2} (left panel) shows curves from Eq.~(\ref{nperpeq}). To illustrate the acceptance effect of the trigger condition the dotted curve omits both $g$ factors, the dashed curve includes $g_s$ and the solid curve includes both factors. The lower hatched band indicates that only a fraction $\epsilon_s$ of the soft component passes the $p_t$ acceptance. The corresponding density for the full soft component would be $2.5/ 2\pi \approx 0.4$, larger than the total $N_\perp$ density $\approx 0.3$ predicted for triggered hard events. The upper hatched band is dominated by the minijet (most-probable jet) contribution to hard events. While the most-probable event multiplicity for hard events is 5.7 the soft component is $5.7 - 2.5 = 3.2$ just as for soft events. Thus, a $p_t$ trigger that selects hard events {\em does not necessarily change the soft-component multiplicity}.
Corresponding 0.9 TeV CMS data are presented in Fig.~3 (upper left) of Ref.~\cite{under}. There is reasonable agreement between the forms but a significant difference in the plateau values. For a leading-track  (single-particle) trigger the plateau begins near $p_{t,trig} = 2.5$ GeV/c, whereas for a leading-track-jet  trigger the plateau begins near $P_j = 5$ GeV/c.

\subsection{Comparison with other collision energies}

Figure~\ref{trends2} (right panel) compares the 200 GeV prediction (solid curve, lower limit) with plateau values from higher collision energies. We consider the CDF result at 1.8 TeV in more detail based on 2D angular densities. The same acceptance $p_t > 0.5$ GeV/c and $|\eta| < 1$ is applied. For 200 GeV data we have
soft component $d^2N_{\perp,s} / d\eta d\phi = 5 \times 0.25 / 2 \times 2\pi = 0.1$ or $N_{\perp,s} = 5 \times 0.25 / 3 \approx 0.4$
and hard component $d^2N_{\perp,h} / d\eta d\phi = 2.25  / 2 \times 2\pi \approx 0.18$ or $N_{\perp,h} = 2.25/3 = 0.75$. 
The latter assumes 2.25/3 jet fragments averaged {\em within the TR}. 
For 1.8 TeV data we have
soft component $d^2N_{\perp,s} / d\eta d\phi = 10 \times 0.25 /2 \times 2\pi = 0.2$ or $N_{\perp,s} = 10 \times 0.25 / 3 \approx 0.8$ 
and hard component $d^2N_{\perp,h} / d\eta d\phi = 4.4  / 2 \times 2\pi \approx 0.35$ or $N_{\perp,h} 4.4 / 3 = 1.5$. The estimated relatation between 1.8 TeV and 200 GeV {\em within the TR} is about 2:1.



The calculated 200 GeV TR jet-fragment yield $N_{\perp,h}$ appears to be roughly 50\% of the 1.8 TeV yield. However, a simulation indicates that for the $\Delta \eta = 1$ acceptance in Ref.~\cite{ppprd} the mean fragment inefficiency (acceptance-edge loss) for minimum-bias jets is about 50\%, whereas the inefficiency for $\Delta \eta  =2$ is about 25\%. Addition of particles outside the $\eta$ acceptance but inside the trigger-jet cone as in the CDF analysis should not affect dijet (hard) contributions to the TR.  The imposed particle or jet $p_t$ trigger condition may substantially bias (increase) the jet fragment multiplicity compared to the untriggered 200 GeV study and may bias the soft component to slightly higher density (e.g., $dn_s/d\eta$ = 5 vs 4). Thus, the 200 GeV solid curves in Fig.~\ref{trends2} represent a lower limit. The corrected plateau value may be substantially higher (illustrated by the hatched region above the solid curve), possibly comparable to the 0.9  TeV result.

\section{Glauber model, $\bf p$QCD and jets} \label{gglauber}

From Sec.~\ref{transverse} we conclude that imposition of a $p_{t,trig}$ condition may select for hard events with at least one jet but does not control the soft component, and therefore (according to Ref.~\cite{ppcent2}) should not significantly affect the \pp collision centrality. Most triggered hard events correspond to small $n_{ch}$ and possibly large impact parameter. In Ref.~\cite{ppprd} it was observe that imposition of an event-multiplicity $n_{ch}$ condition controls the jet production rate over a large range, and therefore {\em may} control \pp collision centrality to some extent.  We next  explore possible relations among jet production, imposed $n_{ch}$ condition and \pp centrality in terms of the Glauber model of \aa geometry and pQCD theory.

\subsection{The Glauber model in \aa collisions}

The Glauber model applied to \aa collisions is based on a model nucleon density $\rho_A(r)$, a 3D nuclear density function integrating to A nucleons~\cite{glauber}. The nuclear {\em profile function} is $T_A(\vec s) = \int dz\, \rho_A(z,\vec s)$, the nucleon areal density on a plane spanned by $\vec s$ and perpendicular to chord $z$ through the nucleus. The nuclear {\em overlap function} for intersecting nuclei A and B is 
\bea \label{tab}
T_{AB}(b) &=& \int d^2s\, T_A(\vec s) T_B(\vec b - \vec s), 
\eea
a nucleon-pair density on the plane spanned by A-B impact vector $\vec b$, with $\int d^2b\, T_{AB}(b) = AB$.
The number of \nn binary collisions with cross section $\sigma_{NN}$ is 
\bea
 N_{bin}(b) &\equiv& \sigma_{NN} T_{AB}(b),
\eea
consistent with basic assumptions of the Glauber model: eikonal approximation, straight-line trajectories, linear superposition of \nn collisions. 
The number of participant-nucleon pairs in an \aa collision is
\bea \label{aanpart}
N_{part}(b)/2 &=&  \int d^2s T_A(\vec s) [1 - P_0(\vec b - \vec s)] \leq A,
\eea
where $P_0(\vec b - \vec s) \approx \exp[- \sigma_{NN} T_A(\vec b -\vec s)]$ is the probability that a projectile nucleon does not interact in the partner nucleus.

In 200 GeV \auau collisions hadron production is observed to vary with centrality approximately as~\cite{jetspec}
\bea
n_{ch} &\approx & n_{pp}(0.9 N_{part}/2 + 0.1 N_{bin}) \hskip .1in \text{central Au-Au} \\ \nonumber
&\approx & n_{pp}(0.96 N_{part}/2 + 0.04 N_{bin}) \hskip .1in \text{peripheral Au-Au}
\eea
for $\Delta \eta = 2$. In more-peripheral \aa collisions with \nn linear superposition~\cite{anomalous} $n_{ch}(b) \propto N_{part}(b)$ and jet production scales as $n_j \propto N_{bin}(b) \propto n_{ch}^{4/3}(b)$, also implying that mean participant pathlength $\nu \equiv 2N_{bin} / N_{part} \propto n_{ch}^{1/3}$~\cite{centrality}.  Those trends in \auau can be compared with particle- and jet-production systematics in \pp collisions where fluctuations may dominate mean values.


\subsection{A Glauber model for p-p collisions and pQCD} \label{ppglaub}

A similar approach can be applied (formally at least) to \pp collisions such that projectile nucleons are treated as transverse distributions of partons.
Parton distribution functions (PDFs) such as $f_g(x|Q^2)$ (for gluons) are expressed in terms of parton momentum space relative to the hadron momentum. To model possible effects of \pp centrality we require a differential parton distribution function that represents the transverse geometry of hadrons. 
Gluon distribution function $f_g(x,t|Q^2)$ is inferred from DIS measurements ($t$ is  momentum transfer to the target proton, $x$ is the parton momentum fraction within the proton, $Q$ is the parton energy scale). The Fourier transform $f_g(x,s|Q^2)$ is the inferred gluon transverse density within a proton, with $ \int d^2s f_g(x,s|Q^2) = f_g(x|Q^2)$. 
Gluon density $f_g(x,s|Q^2)$ is analogous to nucleon density $T_A(s)$. 

A \pp  overlap function can be defined for given $Q^2$
\bea \label{p2}
P_2(b) \,  f_g(x_1) f_g(x_2) \hspace{-.04in}  &\approx& \hspace{-.04in}  \int d^2s\, f_g(x_1,\vec s) f_g(x_2,\vec b - \vec s) 
\eea
with $ \int d^2b P_2(b) \approx 1$ analogous to $T_{AB}(b)$ defined in Eq.~(\ref{tab}).  The definition assumes an eikonal model with straight-line parton trajectories and approximate factorization of $x$ and $s$ trends within a limited $x$ interval corresponding to low-energy jets (gluons) near mid-rapidity. It also assumes nonfluctuating $f_s(x,s|Q^2)$. Functions $f_g(s)$ and unit-normal $P_2(b)$ are assumed to be represented by Gaussians with widths derived from DIS data~\cite{ppcent2}.


The pQCD dijet spectrum for 2-2 parton (gluon) scattering $a, b \rightarrow c,d$ (collinear factorization) is
\bea \label{qspec}
\frac{d\sigma_{cd}(Q^2)}{dQ^2} \hspace{-.005in} &\propto& \hspace{-.005in} \sum_{a,b} \int dx_1dx_2 f_a(x_1,Q^2) f_b(x_2,Q^2)\\ \nonumber
&\times& d\hat\sigma_{ab \rightarrow cd}(x_1,x_2,Q^2)/dQ^2
\eea
averaged over the distribution of \pp impact parameters. The integral over the dijet spectrum at 200 GeV (with spectrum lower bound at 3 GeV) is $\sigma_{dijet} \approx 2.5$ mb~\cite{fragevo}. We then generalize to $s$-{\em dependent} PDFs $f_g(x,s|Q^2)$ and obtain for a \pp centrality class with impact parameter $b$
\bea \label{sigmab}
\frac{d^2\sigma_{cd}(b,Q^2)}{d^2b\, dQ^2}  \hspace{-.05in} &\propto& \hspace{-.08in}\sum_{a,b} \int dx_1 dx_2 P_2(x_1,x_2,b|Q^2)  \\ \nonumber
&\times&d\hat\sigma_{ab \rightarrow cd}(x_1,x_2,Q^2)/dQ^2 \\ \nonumber
&\approx&  P_2(b)\, \frac{d\sigma_{cd}(Q^2)}{dQ^2}
\eea
analogous to $\sigma_{NN} T_{AB}(b) = N_{bin}(b)$. The integral over $d^2b$ recovers Eq.~(\ref{qspec}). As noted, the second line assumes factorization of PDFs within a limited $x$ interval.

\vskip .1in

In the Glauber context $n_{ch}$ (dominated by the soft component from projectile-nucleon dissociation) may serve as a measure of the number of participant partons $N_{part}$, and observed $n_j$ (dijet number) is a measure of the number of parton binary collisions $N_{bin}$ represented by Eqs.~(\ref{qspec}) and (\ref{sigmab}). The question remains whether a Glauber description of jet production in \pp collisions is relevant to observations. 



\subsection{Measured jet production vs $\bf n_{ch}$ in p-p collisions} \label{nj}


The observed number of minimum-bias (untriggered) dijets per 200 GeV NSD \pp collision within acceptance $\Delta \eta = 2$ is~\cite{fragevo,jetspec}
\bea \label{nj2}
n_j &=&  \frac{\sigma_{dijet}}{\sigma_{NSD}}   \frac{\Delta \eta}{\Delta \eta_{4\pi}}  
\approx 0.03,
\eea
where $\Delta \eta_{4\pi}$ ($\approx 5$ for 200 GeV) is the effective $4\pi$ interval for a dijet distribution uniform on $\eta$, $\sigma_{dijet} \approx 2.5$ mb is the integral of Eq.~(\ref{qspec})~\cite{fragevo} and $\sigma_{NSD} \approx 36$ mb at 200 GeV~\cite{nsd}. The NSD result corresponds to an average over \pp $n_{ch}$ (or centrality). If an $n_{ch}$ condition is imposed on \pp events a \pp centrality dependence may be relevant. 



\begin{figure}[h]
  \includegraphics[width=1.65in,height=1.6in]{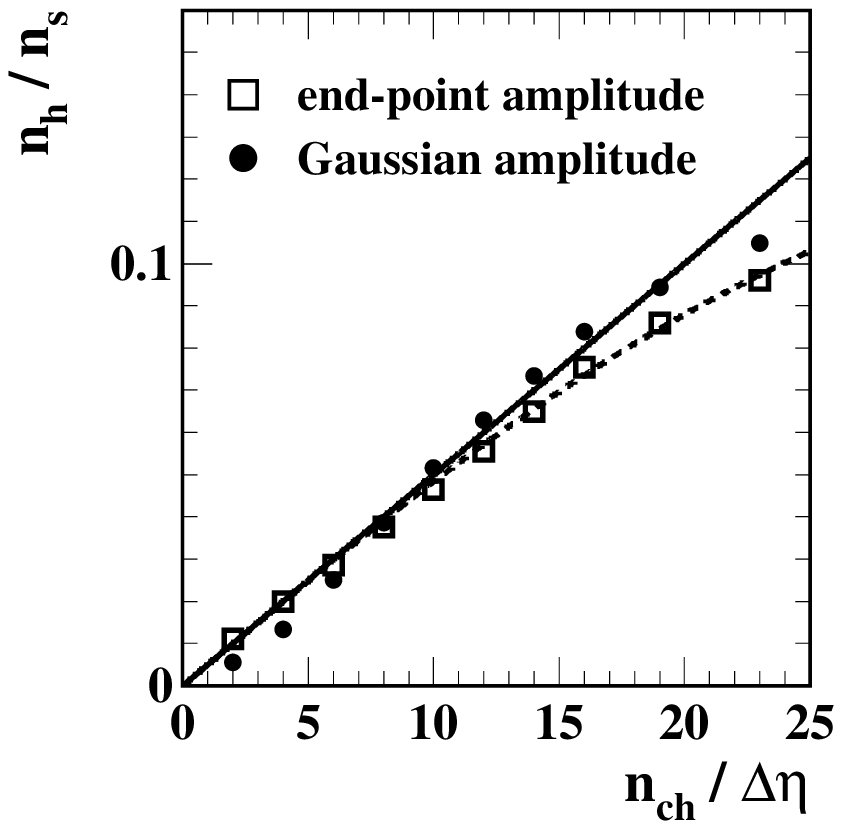}
  \includegraphics[width=1.65in,height=1.6in]{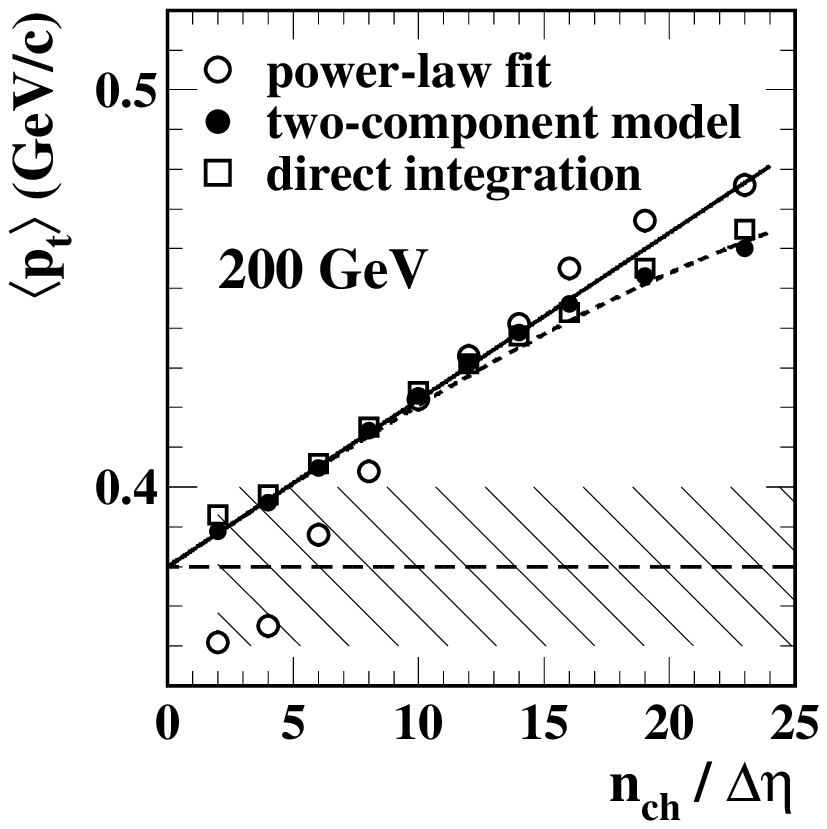}
\caption{\label{jettrends}
Left: 
Hard-component/soft-component multiplicity ratio $n_h / n_s$ vs $n_{ch}$ event class for two analysis methods (points). The solid curve represents the two-component trend $n_h/n_s \approx 0.005\, n_{ch}/\Delta \eta$ equivalent to that reported in Ref.~\cite{ppprd}. The dashed curve estimates possible saturation of the data trend.
Right: 
Ensemble-mean $p_t$ vs $n_{ch}$ event class for three analysis methods (points). The curves (described in the text) are equivalent to those in the left panel.
}  
 \end{figure}

Fig.~\ref{jettrends} (left panel) shows the spectrum hard-component multiplicity $n_h$ (measured within $\Delta \eta = 1$) in the form of ratio $n_h/n_s$ vs $n_{ch}$.
The ratio was estimated by two methods, as noted in the panel and described in Ref.~\cite{ppprd}.
The observed trend is  $n_h/n_s \approx 0.005\, n_{ch}/\Delta \eta$ (solid line)~\cite{ppprd}. That result is consistent within experimental uncertainties with Eq.~(\ref{nj2}) for NSD collisions assuming that the mean jet fragment multiplicity is $\bar n_{ch,j} \approx 2.5$ but with $\Delta \eta = 2$. The corresponding dijet production vs $n_{ch}$ condition is then given by
\bea \label{qnch}
n_j(n_{ch}) 
&\approx& 0.03\,  \frac{(n_{ch} / n_{ch,NSD})^2}{1 + (n_{ch} / n_{ch,0})^2},
\eea
where $n_{ch,NSD}/\Delta\eta = 2.5$ is assumed at 200 GeV, and $n_{ch,0} \gg n_{ch,NSD}$ represents the possibility that the increasing trend may saturate at some point (dashed curve). The number of {\em jets} within $\Delta \eta$ is larger by a factor representing the fraction of dijets with recoil partner in the $\eta$ acceptance, which for $\Delta \eta/\Delta \eta_{4\pi} = 0.4$ is about 1.25~\cite{jetspec}.
Assuming events are Poisson distributed according to dijet multiplicity the  probability of soft events (no dijet in $\Delta \eta = 2$) is $P_0(n_{ch}) = \exp(-n_j)$, the probability of hard events (at least one dijet in $\Delta \eta = 2$) is $1 - P_0(n_{ch})$ and the probability of a {\em second} dijet (MPI) is $P_2(n_{ch}) = [n_j^2(n_{ch})/2] P_0(n_{ch})$ (MPI or {\em multiple parton interactions} is discussed further in Sec.~\ref{mmpi}).

Fig.~\ref{jettrends} (right panel) shows the $\langle p_t \rangle$ vs event multiplicity condition for three spectrum analysis methods described in Ref.~\cite{ppprd}. The ``power law'' $p_t$ spectrum model does not describe the spectra well and introduces significant systematic errors in inferred $\langle p_t \rangle$. The two-component spectrum model effectively extrapolates the spectrum down to $p_t = 0$.  In the TCM the fixed mean $p_t$ for spectrum soft component $S_0(y_t)$ is  $\langle p_{t} \rangle_s = 0.38$ GeV/c (dashed line). The fixed mean $p_t$ for hard component $H_0(y_t)$ is $\langle p_{t} \rangle_h \approx 1.2$ GeV/c. The trend linear on $n_{ch}$ (solid line) is then $(n_s \langle p_{t} \rangle_s + n_h \langle p_{t} \rangle_h ) / n_{ch}$ as described in Ref.~\cite{ppprd}, with $n_s = n_{ch} - n_h$ and $n_h/n_s $ as in the left panel. The $n_h/n_{s}$ and $\langle p_t \rangle$ trends are consistent with minimum-bias jets (minijets) as the underlying mechanism for the $n_{ch}$ dependence. Jet production as inferred from spectrum and correlation data appears to scale as $n_j \propto n_{ch}^2 \approx n_s^2$ over an extended $n_{ch}$ interval. The dashed curve indicates possible saturation of the $\langle p_t \rangle$ trend for larger $n_{ch}$, consistent with the left panel.

Equation~(\ref{nj2}) includes the integral $\sigma_{dijet}$ of the parton spectrum from Eq.~(\ref{qspec}).  Equation~(\ref{qnch}) based on an $n_{ch}$ condition {\em may} correspond to $d\sigma_{dijet}/ d^2b$ (for some $b$ interval) from Eq.~(\ref{sigmab}) assuming $n_{ch}$ is strongly correlated with impact parameter $b$. If such a correlation exists parton participant number $N_{part} \sim n_{ch}$ represents Eq.~(\ref{aanpart}) applied to $f_g(x,s|Q^2)$. The corresponding number of dijets predicted by Eq.~(\ref{sigmab}) (integrated over $Q^2$) should be $n_j \propto N_{bin} \sim n_{ch}^{4/3}$ according to the eikonal assumption adopted by analogy with \aa collisions, for which each participant can only interact with target participants along a straight-line trajectory.

\subsection{Is dijet production related to p-p centrality?} \label{pppcent}

Is a Glauber model of \pp collisions characterized by impact parameter $b$, as in Eq.~(\ref{sigmab}), consistent with observations? We compare $s$-dependent PDFs from a \pp Glauber model in Sec.~\ref{ppglaub} to the observed $n_{ch}$ dependence of untriggered jet production in Sec.~\ref{nj}. We assume that soft-component multiplicity $n_s$ ($\approx n_{ch}$ for lower \pp multiplicities) estimates the number of participant partons [local parton-hadron duality (LPHD)~\cite{lphd}]. The number of participants {\em may be} correlated with \pp centrality. The number of parton-parton binary encounters is estimated by (proportional to) the observed number of dijets  $n_j$.
According to Ref.~\cite{ppcent2} hard processes should arise in more-central \pp collisions. Jet production (probability per event) should then be correlated with \pp centrality. Is such a system of $n_{ch}$, jets, parton participants, binary collisions, Glauber model and DIS data self consistent?


In general, two extremes are possible: (a) mean-value $s$-differential PDFs derived from DIS data remain fixed for all \pp collisions with a given centrality, but different \pp centralities (possibly selected by a jet $p_t$ trigger) lead to different $n_j \propto N_{bin}$ and $n_{ch}$ numbers per Eq.~(\ref{sigmab}), or (b) PDFs fluctuate over a large range, and larger $n_{ch}$ selects for larger $N_{part}$ and $N_{bin}$ with no relation to \pp collision centrality. If we accept theory scenario (a) and the eikonal assumption we expect 
$n_j \propto N_{bin} \propto n_{ch}^{4/3}$ per Eq.~(\ref{sigmab}). If scenario (b) is correct $n_j \propto n_{ch}^2$ should result per Eq.~(\ref{qspec}). If $n_{ch}$ is not significantly correlated with $b$ the system is effectively averaged over $b$ by fluctuations. A participant may interact with any participant from the collision partner and the $n_j \propto n_{ch}^2$ trend follows. 

\begin{figure}[h]
  \includegraphics[width=3.3in,height=1.6in]{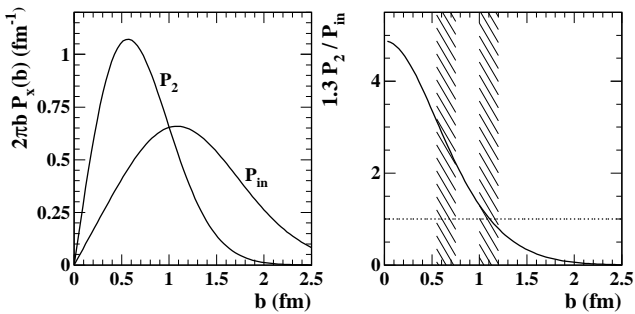}
\caption{\label{theory}
Left: Probability distributions on \pp impact parameter $b$ from Ref.~\cite{ppcent2} for dijet production $P_2$ and \pp inelastic scattering $P_{in}$.
Right: Ratio $P_2/P_{in}$ conjectured to be proportional to the number of dijets per inelastic collision.
}  
 \end{figure}

Figure~\ref{theory} represents eikonal scenario (a) (with mean-value PDFs, no fluctuations) represented by
\bea \label{njtheory}
n_j(b) &=& \frac{\sigma_{dijet}(b)}{\sigma_{pp}(b)} \frac{\Delta \eta}{\Delta \eta_{4\pi}}  \\ \nonumber
&\approx& 0.03 \times 1.3 \frac{ P_2(b)}{P_{in}(b)},
\eea
where $\sigma_{pp}(b) = P_{in}(b)\, \sigma_{NSD}$. In the second line the ratio is essentially $N_{bin}(b) \approx N_{part}^{4/3}(b)$ consistent with the eikonal assumption, and $N_{part}$ corresponds to fixed mean-value PDFs. Figure~\ref{theory} (left panel) shows functions $P_2(b)$ and $P_{in}(b)$ from Ref.~\cite{ppcent2} (at 7 TeV). Figure~\ref{theory} (right panel) shows the ratio factor appearing in Eq.~(\ref{njtheory}). The factor 1.3 brings the ratio factor to unity for $b \approx 1.15$ fm corresponding to NSD \pp collisions (right band).  According to assumptions a jet trigger should select collisions corresponding to $b \approx 0.65$ fm (left band) with an increase in dijet production by a factor less than 3.

Fluctuation-dominated scenario (b) is represented by
\bea
n_j(n_{ch}) &=& \frac{\sigma_{dijet}(n_{ch})}{\sigma_{pp}(n_{ch})} \frac{\Delta \eta}{\Delta \eta_{4\pi}}  \\ \nonumber
&\approx& 0.03 \frac{P_2(n_{ch})}{P_{in}(n_{ch})} \\ \nonumber
&\approx& 0.03\, (n_{ch}/2.5)^2.
\eea
We observe that $P_2(n_{ch}) \approx n_{ch}^2\, P_{in}(n_{ch})$. In that case $N_{bin} \approx N_{part}^2 \rightarrow n_{ch}^2$ consistent with an average over $b$, since $N_{bin} \propto N_{part}^2$ implies that a parton participant may interact with any target participant. The observed $n_j(n_{ch})$ trend appears to falsify the eikonal scenario.

Absolute magnitudes tend to corroborate that conclusion. In \auau collisions $N_{bin}(b) \propto N_{part}^{4/3}$ proportional to dijet production varies with centrality approximately over the interval [1,1000]~\cite{centrality}. Figure~\ref{jettrends} (right panel) shows that for eikonal scenario (a) $N_{bin}(b)$ varies over an interval proportional to [1,3] and cannot describe the observed dijet production $n_j(n_{ch})$ proportional to a range greater than [1,100] for an achievable range [1,10] of $n_{ch}/\Delta \eta$ at 200 GeV. The analogy with \aa collisions can be expressed symbolically by
\bea
n_j \propto N_{bin}(N_{part}) P(N_{part})
\eea
In \auau collisions with fixed $b$ (or $n_{ch}$) $n_{ch} \propto N_{part}(b)$, $N_{bin} \approx N_{part}^{4/3}$ and $P(N_{part}) \sim O(1)$ for a small range of $N_{part}(b)$ values (fluctuations are negligible). In \pp collisions $n_{ch} \approx N_{part}$ but $P(N_{part})$ is broadly distributed and $N_{bin} \propto N_{part}^2$ (fluctuations dominate).

According to Ref.~\cite{ppcent2} by trigging on jets we should select more-central \pp collisions with substantially greater multiplicities: $N_h \gg N_s$ in Eq.~(\ref{nperptheory}). We do observe that multiplicity and jet production are highly correlated and vary over a large range.  If the proton PDF does not fluctuate where does the large $n_{ch}$ in some \pp events come from? If $n_{ch}$ is strongly correlated with $b$ then large $n_{ch}$ should correspond to {\em very} central \pp collisions. If selecting dijet events biases to central collisions  then a jet trigger should result in greatly increased $n_{ch}$.  But the $n_{ch}$ increase produced by a jet trigger is only a small fraction of the available $n_{ch}$ range and should correspond to peripheral \pp collisions according to the eikonal scenario. 


From the simulation exercise in Sec.~\ref{ttotal} we conclude that it is much easier to bias to hard events with a small $n_{ch}$ increase than with a substantial decrease in $b$. From the presently-available data we cannot determine the degree of correlation between $n_{ch}$ and $b$. We cannot rule out some degree of correlation for the lowest multiplicities. For larger multiplicities correlation seems to be negligible and fluctuations dominate. Extension to analysis of correlation data vs $n_{ch}$ may better access that relation. 



 \section{Discussion} \label{disc}

This study addresses the relations among $p_t$ trigger conditions, event multiplicity $n_{ch}$ conditions, dijet production and \pp collision centrality. What can be inferred from existing data about correlations between $p_{t,trig}$ and $b$, between $n_{ch}$ and $b$, and what is the composition of the underlying event? We reconsider the physical mechanisms responsible for observed $N_\perp$ vs $p_{t,trig}$ trends and $N_\perp(p_t)$ spectra. We introduce the concept of a universal dijet base that is broad on azimuth and common to all jet angular correlations. We conclude that multiple parton interactions are unlikely for imposed $p_{t,trig}$ conditions but may occur with large event multiplicities.

\subsection{Comparing $\bf N_\perp$ spectra at different energies} \label{lowspec}

The collision-energy dependence of $N_\perp$ systematics, including the $N_\perp(p_t)$ spectrum, is of considerable interest for UE studies. To establish such a trend systematic uncertainties and correspondence of analysis conditions must be established. A simulation carried out in Secs.~\ref{ttotal} and \ref{transverse} based on 200 GeV \pp spectrum and correlation data is compared to measurements from 0.9 and 1.8 TeV. The simulation provides insight into UE comparisons.

In Fig.~\ref{spectrum} the $N_\perp$ spectrum for 0.9 TeV from Fig. 4 (bottom) of Ref.~\cite{under} (open points) is compared with the spectrum for 1.8 TeV from Fig.~37 of Ref.~\cite{under5} (solid points).  The $\eta$ acceptances differs---$\Delta \eta = 4$ at 0.9 TeV vs $\Delta \eta  =2$ at 1.8 TeV---and the jet trigger condition is substantially different---$P_j > 5$ GeV/c at 1.8 TeV vs $P_j > 3$ GeV/c at 0.9 TeV. Such differences can complicate interpretation of apparent energy dependence. The two-component model can aid such comparisons. 

The soft-component contribution $N_{\perp,s}(p_t)$ at 0.9 TeV should be a factor $2 \times 3.5/4 \approx 1.8$ times larger than that at 1.8 TeV, reflecting differences in $\eta$ acceptance and energy-dependent hadron $\eta$ density. The ratio 1.8 is consistent with the lowest data points (at 0.75 GeV/c) in Fig.~\ref{spectrum}. Since the accepted soft component is approximately one third of $N_\perp$ that is a significant issue.

The hard component $N_{\perp,h}(p_t)$ is influenced by both $\eta$ acceptance and jet trigger condition. The two-times larger $\eta$ acceptance at 0.9 TeV should lead to increased jet acceptance, but the different trigger condition goes in the opposite direction.  The higher trigger at 1.8 TeV achieves the plateau value 2.3 in Fig.~30 of Ref.~\cite{under5}, whereas the lower trigger at 0.9 TeV achieves only 50\% of the plateau value in Fig.~3 (right) of Ref.~\cite{under}. The combination may produce comparable hard-component yields but a softer  hard spectrum shape at 0.9 TeV. 

 
\subsection{High-$\bf p_t$ trigger condition $\bf vs$ $\bf p$-$\bf p$ centrality}

The trends in Fig.~\ref{trends2} and similar results from UE studies such as Ref.~\cite{under5} can be misinterpreted. According to assumptions $N_\perp$ is a ``soft'' multiplicity not associated with the triggered dijet, its variation appearing to match the scenario described in Ref.~\cite{ppcent2}. However, $N_\perp$ actually includes both soft (nonjet) and hard (trigger-jet-related) components. 
In Sec.~\ref{etype} we demonstrate that while imposition of a high-$p_t$ trigger condition does bias toward hard events (as described by the two-component model) it does not significantly increase the soft component of $N_\perp$ in accepted \pp events, which remains systematically consistent with the NSD value. The fraction {\em accepted} by the effective upper $p_{t,trig}$ cutoff does increase from zero at $p_{t,min}$ up a soft plateau value at about 1 GeV/c.

Given the conjecture in Ref.~\cite{ppcent2} the inferred soft-component trend implies that the $p_t$ trigger does not significantly alter the collision centrality. The {\em event-wise} jet production probability in peripheral events may be small (few percent) but the corresponding fraction of all events is large enough (Fig.~\ref{ppmult}) that small $N_\perp$ (and possibly large $b$) still dominates the overall jet production process.

The hard-event fraction does increase with $p_{t,trig}$ until the jet frequency per event rises to unity near $p_{t,trig} = 2.5$ GeV/c as described by the TCM. Further increase of $p_{t,trig}$ should bias to higher dijet energy, but the jet-related structure (dijet base) {\em within the TR} should be insensitive to subsequent small-angle jet growth. The TCM as applied in the present study with its simplified model of the hard component is not intended to describe $N_\perp(p_t)$ and $N_\perp(p_{t,trig})$ trends above about 6 GeV/c.






\subsection{Event multiplicity condition vs $\bf p$-$\bf p$ centrality}

The relation between $n_{ch}$ and \pp centrality remains unclear. In \auau collisions $n_{ch}$ and $b$ are strongly correlated. Fluctuations of geometry and particle production play a minor role. 
In Ref.~\cite{ppprd} it was established that imposing a condition on $n_{ch}$ in \pp collisions has negligible influence on jet properties but directly controls the event-wise frequency of minimum-bias dijets. 
%
However, in \pp collisions fluctuations may dominate mean values, eliminating any significant correlation between $n_{ch}$ and $b$ except perhaps for peripheral collisions.
%
%
The low-$x$ parton structure of the nucleon relevant to both minimum-bias jets and projectile dissociation may fluctuate strongly. 
Both participant-parton number ($\propto n_{ch}$) and parton binary collisions ($\propto n_j$) would then fluctuate  with some broad frequency distribution unrelated to centrality.




We {observe} that the two quantities are strongly correlated ($n_j \propto n_{ch}^2$). Given the form of Eq.~(\ref{sigmab}) that relation could arise because the eikonal approximation is not valid and all participants can interact or because $b$ is effectively averaged over a large interval and the system reverts to Eq.~(\ref{qspec}), albeit with a large increase in the participant number due to fluctuations of the proton structure. 
In the latter case the form of the minimum-bias distribution for larger $n_{ch}$ would not relate to the \pp impact parameter, only to fluctuations in the proton low-$x$ structure. We conclude that $b$ and $n_{ch}$ might be significantly correlated at smaller $n_{ch}$  but may be uncorrelated at larger $n_{ch}$. The average $b$ may then retain a relatively large value nearly independent of $n_{ch}$.

\subsection{Composition of the Underlying Event}




The event components commonly assumed in analysis of the underlying event are (a) an energetic dijet, (b)  a beam-beam (soft) component representing projectile dissociation, (c) initial- and final-state QCD radiation, (d) possible multiple parton interactions (MPI, additional minimum-bias hard scatters)~\cite{rick,under5,under}. The UE is identified as the combination (b) + (c) + (d). The {transverse region} on azimuth with multiplicity $N_\perp$ is said to be particularly sensitive to the UE: the hard dijet (a) should not contribute significantly to the TR. Extrapolation of the $N_\perp$ $p_t$ spectrum to zero seems to indicate a UE particle yield $2.5\times$ larger than the soft component, suggesting a novel production mechanism such as MPI.

Some of those assumptions can be questioned given measured (a) jet fragmentation functions and (b) jet angular correlations: (a) As shown in Fig.~3 (left panel) of Ref.~\cite{eeprd}, with increasing jet energy the fragment density (fragmentation function) on rapidity $y = \ln[(E + p)/m_\pi]$ 
increases at larger $y$ in a self-similar manner. Higher-energy jets are built on lower-energy jets by adding  more higher-momentum fragments which  appear at {\em smaller angles} relative to the dijet (thrust) axis.  (b) Structure along the dijet axis  ($y_z$) can be described as a central ``string-like'' region terminated by ``limiting-fragmentation'' regions at the ends. The central region on $y_z$, its particle density slowly-varying with jet energy,  corresponds to {\em large} relative angles that overlap the TR.


In Sec.~\ref{angcorr} measured angular correlations for minimum-bias jets reveal that  no azimuth region excludes jet structure.  While the 2D distribution is highly structured and the SS and AS jet components can be clearly distinguished,  the projection of {\em minimum-bias} jet correlations onto 1D azimuth is almost uniform over $2\pi$, as demonstrated in Fig.~\ref{cdf}. The jet contribution to the TR is then substantial. As the triggered jet momentum or $P_j$ increases additional particles at larger $p_t$ should appear at smaller angles, but the trigger-jet contribution to the TR remains the same as for minimum-bias jets. 

Given the results in Sec.~\ref{angcorr} the region $|\phi_\Delta| < 0.5$ and $|\eta_\Delta| > 1$ would be a better choice to study the UE. It is well outside both the SS and AS ``jet cones'' and should exclude most of the central region (on $y_z$) of any dijet. Reconstruction of 2D angular correlations relative to the trigger-jet axis in future UE studies might provide improved (more differential) access to the {\em nonjet} UE.

\subsection{Relevance of multiple parton interactions} \label{mmpi}


Recent developments in this topic are reviewed in Ref.~\cite{mpi}. Multiple parton interactions have been expected to form a major fraction of the UE~\cite{under}. The hard component of $N_\perp$ inferred from the present study might then represent the MPI contribution. MPI are generated in Monte Carlos such as \textsc{pythia}~\cite{smpi} and  \textsc{herwig}~\cite{hmpi} conventionally used to model UE systematics. A critical aspect of such MCs is the scattered-parton spectrum cutoff energy which is treated as an adjustable parameter. Significant MPI rates are achieved by lowering the cutoff energy so that the effective 2-2 parton cross section is comparable to the \nn  cross section for a given \pp event trigger (e.g., NSD collisions, $p_{t,trig}$ trigger condition).

A parton spectrum lower bound can be inferred directly from \pp and \auau spectrum and correlation data~\cite{fragevo,jetspec}. The data in Fig.~\ref{ppspec} from Ref.~\cite{ppprd} indicate that the minimum-bias jet frequency for NSD events  is consistent with a 2.5 mb parton cross section (spectrum integral). And the pQCD description in the right panel of that figure requires a parton spectrum cutoff near 3 GeV consistent with $\sigma_{dijet} \approx $ 2.5 mb.
The present study indicates that imposition of a $p_t$ trigger condition does not significantly bias the soft multiplicity or the collision centrality. The probability of a second, minimum-bias or semihard dijet accompanying a triggered hard dijet should be only a few percent, as for untriggered events. 

If an $n_{ch}$ condition is imposed the situation changes dramatically since $n_j \propto n_{ch}^2$~\cite{ppprd}. From Eq.~(\ref{qnch}) $n_{ch}/\Delta \eta \approx 13$ implies $n_j \approx 1$, and the probability of at least one MPI is $\approx 35$\%. From Fig.~\ref{ppspec} (left panel) we observe that the hard component (fragment distribution) does not change appreciably with $n_{ch}$, suggesting that the underlying parton spectrum and 2-2 cross section remain unchanged. From the argument in Sec.~\ref{pppcent} we conclude that the $n_{ch}$ condition biases to larger parton participant number and therefore to multiple parton binary collisions. A $p_{t,trig}$ condition does not significantly alter the (small) probability of MPI in \pp collisions. The observed hard component of $N_\perp$ is a manifestation of the triggered dijet.

\section{Summary} \label{summ}


Triggered dijets produced in high-energy \pp collisions are by definition accompanied by a complementary {\em underlying event} or UE. The UE is assumed to represent several processes, including projectile-proton dissociation, initial- and final-state radiation and possibly other mechanisms such as {\em multiple parton interactions} or MPI. It is further assumed that dijet structure is confined to regions on $(\eta,\phi)$ with $R<1$ at 0 and $\pi$ azimuth. The UE is then studied in a {\em transverse region} or TR centered at $\pi/2$ azimuth. The multiplicity $N_\perp$ integrated within the azimuth TR and some $\Delta \eta$ acceptance is the primary UE observable. The $N_\perp$ $p_t$ spectrum and responses to imposed conditions are employed to study the UE.

UE topics of current interest include verifying the physical mechanisms that contribute to the TR, especially the extent of MPI contributions, and the role of \pp collision centrality. 
The primary observables are $N_\perp(p_{t})$ spectra, triggered $N_\perp(p_{t,trig})$ systematics and dijet production rates.
Measured trends have been compared to Monte Carlo predictions with a variety of parameter tunes. A key issue for Monte Carlos is the invoked pQCD parton spectrum model and consequences for the extent of MPI.


In the present study we introduce a two-component model of minimum-bias and $n_{ch}$-dependent spectrum and correlation data from \pp collisions at 200 GeV to investigate UE composition and $N_\perp$ yield and spectrum trends for several collision energies. We test a theoretical conjecture motivated by DIS data that selection of hard collisions (dijets) with a $p_{t,trig}$ condition favors more-central \pp collisions and increased ``soft'' multiplicities.

A two-component analysis of $N_\perp(p_t)$ spectra reveals that the triggered soft-component yield is not significantly larger than that for minimum-bias collisions, and the soft-component spectrum shape may not change significantly over a range of energies. The common hard-component shape is consistent with a minimum-bias fragment distribution predicted by pQCD, and the yields are quantitatively comparable with a 200 GeV analysis.

We observe that a $p_{t,trig}$ condition can effectively select for hard collisions, those including a (triggered) dijet, but two-component analysis reveals that triggered hard events include ``soft'' hadron densities $dn_{s}/d\eta$ close to the minimum-bias value for a given collision energy. There is no indication that such triggered hard \pp events are ``central'' collisions. 

$N_\perp(p_{t,trig})$ trends are simply explained in the two-component context. $N_\perp$ includes soft and hard contributions. As noted the soft contribution is consistent with (but may be slightly larger than) the measured minimum-bias $dn_{ch}/d\eta$. The hard component is consistent with minimum-bias jet-related angular correlations associated with the triggered dijet itself. The assumption that the TR excludes a triggered dijet contribution is incorrect. 
%
All dijets include a common base corresponding to the central region of the dijet on rapidity along the dijet axis (string). Those fragments emerge at large angular separation from the dijet axis and contribute to the TR.

The observed hard component of $N_\perp$ might be attributed to multiple parton interactions or MPI in more-central \pp collisions. Larger minimum-bias dijet production can be generated in Monte Carlos by reducing the parton spectrum effective lower bound. However, parton spectrum properties are strongly constrained by minimum-bias \pp data. The parton spectrum integral at 200 GeV is approximately 2.5 mb as inferred from the $n_{ch}$ dependence of hadron spectra and consistent with a parton spectrum {\em cutoff} near 3 GeV, implying a minimum-bias dijet probability of a few percent. The incidence of MPI for an imposed $p_{t,trig}$ condition is small but can dominate collisions if a {\em large-$n_{ch}$ condition} is imposed.

In conclusion, the two-component model provides a good description of the UE as well as other aspects of the \pp final state. The status of the theoretical conjecture that mid-rapidity jet production may depend on \pp collision centrality due to the transverse gluon distribution in the proton remains unclear. This exercise based on yields and spectra points to the need for correlation studies, especially the nonjet azimuth quadrupole, that may provide additional information about \pp centrality.


\vskip .1in
This work was supported in part by the Office of Science of the U.S. DOE under grant DE-FG03-97ER41020.

\begin{appendix}

\section{TCM model functions} \label{tcmmodels}

The unit-integral functions for the two-component model (TCM) of $m_t$ or $y_t$ spectra as reported in Refs.~\cite{ppprd,hardspec,fragevo} are defined here. For 200 GeV \pp collisions the soft-component model (L\'evy distribution on $m_t$) is
\bea
S_0(y_t) = \frac{20}{[1 + (m_t - m_\pi)/{nT}]^n}
\eea
with $m_t = m_\pi \cosh(y_t)$, $n = 12.8$ and $T = 0.145$ GeV. The Gaussian form of the hard-component model is
\bea
H_0(y_t) = 0.33 \exp\{-[(y_t - y_{t0})/\sigma_{y_t}]^2/2\}
\eea
on $y_t$, with $y_{t0} = 2.67$ ($p_t \approx 1$ GeV/c) and $\sigma_{y_t} = 0.445$. The coefficients (determined by the unit-integral condition) depend on the specific model parameters. In Ref.~\cite{hardspec} the hard-component model function is generalized to a Gaussian with power-law tail to accommodate the underlying parton energy spectrum. In Ref.~\cite{fragevo} the calculated pQCD hard component derived from fragmentation functions that describes the spectrum data accurately also deviates from a Gaussian below the distribution mode. The simplified Gaussian form adopted for the present analysis is compared with the 200 GeV pQCD calculation and \pp data in Fig.~\ref{ppspec} (right panel). 

\end{appendix}



\begin{thebibliography}{99}

\bibitem{ppcent2}  L.~Frankfurt, M.~Strikman and C.~Weiss,
  Phys.\ Rev.\  D {\bf 83}, 054012 (2011).

\bibitem{ppprd} J.~Adams {\it et al.}  (STAR Collaboration),
  Phys.\ Rev.\  D {\bf 74}, 032006 (2006).

\bibitem{hardspec}  T.~A.~Trainor,
  Int.\ J.\ Mod.\ Phys.\  E {\bf 17}, 1499 (2008).

\bibitem{porter2} R.~J.~Porter and T.~A.~Trainor  (STAR Collaboration),
  J.\ Phys.\ Conf.\ Ser.\  {\bf 27}, 98 (2005).

\bibitem{porter3}  R.~J.~Porter and T.~A.~Trainor  (STAR Collaboration),
  PoS C {\bf FRNC2006}, 004 (2006).

\bibitem{fragevo}    T.~A.~Trainor,
  Phys.\ Rev.\  C {\bf 80}, 044901 (2009).

\bibitem{anomalous}  G.\ Agakishiev, {\it et al.} (STAR Collaboration),
  arXiv:1109.4380.

\bibitem{davidhq}  D.~T.~Kettler  (STAR collaboration),
  Eur.\ Phys.\ J.\  C {\bf 62}, 175 (2009).

\bibitem{davidhq2}  D.~T.~Kettler  (STAR Collaboration),
  arXiv:1008.4793.

\bibitem{under5}  T.~Affolder {\it et al.}  (CDF Collaboration),
  Phys.\ Rev.\ D {\bf 65}, 092002 (2002).

\bibitem{rick} R.~Field,
  Acta Phys.\ Polon.\ B {\bf 42}, 2631 (2011).

\bibitem{under} V.~Khachatryan {\it et al.}  (CMS Collaboration),
  Eur.\ Phys.\ J.\ C {\bf 70}, 555 (2010).

\bibitem{mpi} P.~Bartalini, E.~L.~Berger, B.~Blok, G.~Calucci, R.~Corke, M.~Diehl, Y.~Dokshitzer and L.~Fano {\it et al.},
  arXiv:1111.0469.

\bibitem{kn} D.~Kharzeev and M.~Nardi,
  Phys.\ Lett.\ B {\bf 507}, 121 (2001).

\bibitem{centrality} T.~A.~Trainor and D.~J.~Prindle,
  hep-ph/0411217.

\bibitem{eeprd}  T.~A.~Trainor and D.~T.~Kettler,
  Phys.\ Rev.\ D {\bf 74}, 034012 (2006).

\bibitem{jetspec}  T.~A.~Trainor and D.~T.~Kettler,
  Phys.\ Rev.\ C {\bf 83}, 034903 (2011).

\bibitem{cmsmult} V.~Khachatryan {\it et al.}  (CMS Collaboration),
  JHEP {\bf 1101}, 079 (2011).

\bibitem{kno} Z.~Koba, H.~B.~Nielsen, and P.~Olesen, 
Nucl. Phys. B {\bf 40}, 317 (1972).

\bibitem{tzyam} T.~A.~Trainor,
  Phys.\ Rev.\  C {\bf 81}, 014905 (2010).


\bibitem{glauber} K.~J.~Eskola, R.~Vogt and X.~N.~Wang,
  Int.\ J.\ Mod.\ Phys.\ A {\bf 10}, 3087 (1995).

\bibitem{nsd} G.~J.~Alner {\it et al.}  (UA5 Collaboration),
  Z.\ Phys.\  C {\bf 32}, 153 (1986).

\bibitem{lphd} Ya.~I.~Azimov, Yu.~L.~Dokshitzer, V.~A.~Khoze, S.~I.~Troyan, Z. Phys. C {\bf 27}, 65 (1985),  Z. Phys. C {\bf 31}, 213 (1986).

\bibitem{smpi}  R.~Corke and T.~Sjostrand,
  JHEP {\bf 1103}, 032 (2011).

\bibitem{hmpi} S.~Gieseke, C.~Rohr and A.~Siodmok,
  arXiv:1206.2205.

\end{thebibliography}
\end{document}